\documentclass[twocolumn]{aastex62}

\newcommand{\teff}{$T_{\rm eff}$}
\newcommand{\logg}{$\log$ g}
\newcommand{\vsini}{$v \sin i$}

\submitjournal{ApJ}

\shortauthors{L\'opez-Valdivia et al.}
\shorttitle{}

\usepackage{url}
\usepackage{amsmath}

\begin{document}


\title{The IGRINS YSO Survey I. Stellar parameters of pre-main sequence stars in Taurus-Auriga}


    \author[0000-0002-7795-0018]{Ricardo L\'opez-Valdivia}
\affil{The University of Texas at Austin,
Department of Astronomy, 
2515 Speedway, Stop C1400, 
Austin, TX 78712-1205}
\email{rlopezv@utexas.edu}

\author[0000-0002-3621-1155]{Kimberly R. Sokal}
\affil{The University of Texas at Austin,
Department of Astronomy,
2515 Speedway, Stop C1400,
Austin, TX 78712-1205}

\author[0000-0001-7875-6391]{Gregory N. Mace}
\affil{The University of Texas at Austin,
Department of Astronomy, 
2515 Speedway, Stop C1400,
Austin, TX 78712-1205}

\author{Benjamin T. Kidder}
\affil{The University of Texas at Austin,
Department of Astronomy,
2515 Speedway, Stop C1400,
Austin, TX 78712-1205} 

\author[0000-0001-9580-1043]{Maryam Hussaini}
\affil{The University of Texas at Austin,
Department of Astronomy,
2515 Speedway, Stop C1400,
Austin, TX 78712-1205}

\author{Larissa Nofi}
\affil{Lowell Observatory, 1400 W. Mars Hill Rd, Flagstaff. AZ 86001. USA}
\affil{Institute for Astronomy,  University of Hawaii Manoa, 2680 Woodlawn Drive, 
Honolulu, HI 96822, USA}

\author[0000-0001-7998-226X]{L. Prato}
\affil{Lowell Observatory, 1400 W. Mars Hill Rd, Flagstaff. AZ 86001. USA}

\author[0000-0002-8828-6386]{Christopher M. Johns-Krull}
\affil{Physics \& Astronomy Dept., Rice University,
6100 Main St., Houston, TX 77005}

\author[0000-0002-0418-5335]{Heeyoung Oh}
\affil{The University of Texas at Austin, Department of Astronomy, 2515 Speedway, Stop C1400, Austin, TX 78712-1205}
\affil{Korea Astronomy and Space Science Institute, 776 Daedeok-daero, 
Yuseong-gu, Daejeon 34055, Korea}

\author{Jae-Joon Lee}
\affil{Korea Astronomy and Space Science Institute, 776 Daedeok-daero, 
Yuseong-gu, Daejeon 34055, Korea}

\author{Chan Park} 
\affil{Korea Astronomy and Space Science Institute, 776 Daedeok-daero, Yuseong-gu, Daejeon 34055, Korea}

\author{Jae Sok Oh} 
\affil{Korea Astronomy and Space Science Institute, 776 Daedeok-daero, 
Yuseong-gu, Daejeon 34055, Korea}

\author{Adam Kraus}
\affil{The University of Texas at Austin,
Department of Astronomy,
2515 Speedway, Stop C1400,
Austin, TX  78712-1205}

\author[0000-0001-6909-3856]{Kyle F. Kaplan} 
\affil{SOFIA Science Center - USRA, NASA Ames Research Center, Moffett Field, CA 94035, USA}

\author{Joe Llama}
\affil{Lowell Observatory, 1400 W. Mars Hill Rd, Flagstaff. AZ 86001. USA}

\author[0000-0003-3654-1602]{Andrew W. Mann}
\affil{Department of Physics and Astronomy, University of North Carolina at Chapel Hill, Chapel Hill, NC 27599, USA}

\author[0000-0003-4770-688X]{Hwihyun Kim}
\affil{Gemini Observatory/NSF's NOIRLab, Casilla 603, La Serena, Chile}

\author{Michael A. Gully-Santiago}
\affil{The University of Texas at Austin, Department of Astronomy, 2515 Speedway, Stop C1400, Austin, TX  78712-1205}

\author[0000-0003-0871-3665]{Hye-In Lee} 
\affil{Korea Astronomy and Space Science Institute, 776 Daedeok-daero, 
Yuseong-gu, Daejeon 34055, Korea} 

\author[0000-0002-2548-238X]{Soojong Pak} 
\affil{School of Space Research, Kyung Hee University, 1732 Deogyeong-daero, 
Giheung-gu, Yongin-si,Gyeonggi-do 17104, Korea} 

\author{Narae Hwang}
\affil{Korea Astronomy and Space Science Institute, 776 Daedeok-daero, 
Yuseong-gu, Daejeon 34055, Korea}

\author[0000-0003-3577-3540]{Daniel T. Jaffe}
\affil{The University of Texas at Austin,
Department of Astronomy, 2515 Speedway, Stop C1400, Austin, TX  78712-1205}

\begin{abstract}
We present fundamental parameters for 110 canonical K- \& M-type (1.3$-$0.13~$M_\odot$) Taurus-Auriga young stellar objects (YSOs). The analysis produces a simultaneous determination of effective temperature (\teff), surface gravity (\logg), magnetic field strength (B), and projected rotational velocity (\vsini).  
Our method employed synthetic spectra and high-resolution (R$\sim$45,000) near-infrared spectra taken with the Immersion GRating INfrared Spectrometer (IGRINS) to fit specific K-band spectral regions most sensitive to those parameters. The use of these high-resolution spectra reduces the influence of distance uncertainties, reddening, and non-photospheric continuum emission on the parameter determinations.
The median total (fit + systematic) uncertainties were 170~K, 0.28~dex, 0.60~kG, 2.5~km~s$^{-1}$ for \teff, \logg, B, and \vsini, respectively. 
We determined B for 41 Taurus YSOs (upper limits for the remainder) and find systematic offsets (lower \teff, higher \logg\ and \vsini) in parameters when B is measurable but not considered in the fit. 
The average \logg\ for the Class~{\sc ii} and Class~{\sc iii} objects differs by 0.23$\pm$0.05~dex, which is consistent with Class~{\sc iii} objects being the more evolved members of the star-forming region.
However, the dispersion in \logg\ is greater than the uncertainties, which highlights how the YSO classification correlates with age (\logg), yet there are exceptionally young (lower \logg)  Class~{\sc iii} YSOs and relatively old (higher \logg) Class~{\sc ii} YSOs with unexplained evolutionary histories. The spectra from this work are provided in an online repository along with TW Hydrae Association (TWA) comparison objects and the model grid used in our analysis.

\end{abstract}



\section{Introduction}
Young stellar objects (YSOs) are stars at an early stage of evolution, and the disks surrounding them provide the raw material out of which planets can form.
An empirically determined lifetime for protoplanetary disks is key to differentiating theories of planet formation and migration \citep[e.g., ][]{baruteau14}. At a population level, observations of disk fractions in star forming regions and moving groups, with ages determined by comparing stellar properties to theoretical isochrones, have established a characteristic timescale of 2-3~Myr for YSO disk dissipation \citep[e.g, ][]{haisch01,hernandez07,mamajek09,ribas14}.
These studies also make it clear that there is heterogeneity in the YSO population, with very young clusters having diskless members and a small number of systems with thick disks remaining at 8-10~Myr \citep[e.g. TW Hydra,][and references therein]{sokal18}. There are some indications of differences as a function of mass \citep{ribas15,galli15}, but mass-dependence does not explain many of the exceptional cases.

The combination of ALMA imaging of the outer disks of YSOs \citep[e.g., ][]{alma15,barenfeld17,huang18,long19,perez20} and infrared interferometry and spectroscopy of inner disks \citep[e.g.,][]{muzerolle03,varga17,boccaletti2020} is providing new details on disk structure, kinematics, and composition. 
Both the puzzle of lifetime variation and the new details about the disks themselves drive a need for more precise information about the stellar hosts. 
Effective temperature (\teff) and surface gravity (\logg), as proxies for YSO masses and ages, respectively \citep[see Figure 9 of][]{sokal18}, permit comparisons of disk classifications with evolutionary models \citep{baraffe15, simon19}. Since angular momentum transport is very important both to the disks and to the evolution of stellar rotation { \citep[e.g.,][]{attridge92,herbst07,bouvier14}}, measures of the rotation are also needed. Magnetic interactions between a star and disk are crucial for accretion onto the star and material flow through the disk { \citep[e.g.,][]{camenzind90,koenigl91,lopez16}}. We therefore also need estimates of stellar magnetic field strength.

However, measurements of YSO parameters are influenced by interstellar reddening, continuum veiling, and stellar spots.
The impacts of interstellar reddening are reduced at longer wavelengths, making the infrared more amenable to YSO studies than visible-light observations.  
The presence of strong magnetic fields (B) is associated with cool spots \cite[e.g.,][]{crockett12,gully17}, Zeeman-split atomic spectral lines \citep[e.g.,][]{johns-krull99,doppmann03-2,yang05,yang11,lavail17,sokal18,lavail19,sokal20} and changes in absorption line profiles.
Veiling ($r$) is a non-stellar continuum emission that reduces the depth of the photospheric lines in YSO spectra \citep{joy49}. It has been explained as the consequence of different physical processes including an active chromosphere \citep[e.g.,][]{calvet87,basri89,hartigan89,hartmann90,valenti93,cieza05}, emission produced by accretion onto the star \cite[e.g., ][]{kenyon87,johns-krull01,gahm08}, excess emission from dust at the sublimation radius of the inner disk wall \citep[e.g.,][]{natta01,muzerolle03} and the emission from warm gas located inside the dust sublimation radius of the disk \citep[e.g.,][]{fischer11}.
These variable processes can result in YSO parameters that differ significantly between literature references because the YSOs were observed at different times, in different physical states, and with instruments observing different wavelengths.

We have employed the Immersion GRating INfrared Spectrometer \citep[IGRINS;][]{yuk10,park14,mace16} to survey the Taurus-Auriga star-forming region and determine YSO properties while minimizing the effects listed above.
With its broad spectral grasp (1.45-2.5~$\mu$m) at high resolution (R$\sim$45,000), IGRINS simultaneously observes numerous spectral features of YSOs.
Additionally, we can average over variability and derive typical parameters for each target by combining multiple observation epochs.
In 2014, IGRINS was commissioned on the 2.7~m Harlan J.\ Smith Telescope at McDonald Observatory. Observations initially focused on the Taurus-Auriga complex (often recognized as Taurus) because it is a close (d~=~140~pc, \citealp{kenyon94,galli18}) and young (about 1-5~Myr, \citealp{kraus09,gennaro12}) star-forming region that provides observations of YSOs at various stages of evolution. The population of Taurus is at an age coincident with disk dissipation timescales and includes both Class~{\sc ii} and Class~{\sc iii} YSOs. 

The IGRINS YSO Survey is uniquely suited to the determination of physical parameters because:
\begin{itemize}
    \item[-] The 1.45--2.5~$\mu$m spectral region includes both photospheric and disk contributions to the YSO spectrum.
    \item[-] The fixed spectral format of IGRINS provides similar spectral products for each object at each epoch.
    \item[-] Multiple-epoch observations provide a means to characterize and average over variability.
    \item[-] The spectral resolution of $\sim$7~km~s$^{-1}$ is smaller than the typical rotational velocity of young stars and can reliably resolve line-splitting by magnetic fields (B) $\gtrsim$~1.0 kG.
\end{itemize}

In this paper, we present the simultaneous determination of \teff, \logg, B, and projected rotational 
velocity (\vsini) of 119 K- \& M-type YSOs located in Taurus, of which 110 are reliable.
In Section 2 we describe the sample and the IGRINS observations. Section 3 details the parameter determination and sources of uncertainty, while in Section 4 we discuss the results and provide analysis. In Section 5 we provide a summary and the conclusions of this work.

\section{Observations and sample}
\label{sect:obs}
IGRINS employs a silicon immersion grating as the primary disperser \citep[][]{jaffe98, wang10, gully10} and volume phase holographic gratings to cross-disperse the H- and K-band echellograms onto Teledyne Hawaii-2RG arrays. This setup provides a compact design with high sensitivity and a significant single-exposure spectral grasp \citep{yuk10,park14}. 
IGRINS has a fixed spectral format and no moving optics, so the science products remain unchanged no matter where it is installed.
IGRINS has increased its scientific value by traveling between McDonald Observatory, the Lowell Discovery Telescope (LDT), and the Gemini South telescope \citep{mace18}.

Flat-field calibration frames were taken at the start of each night and used for bad pixel masking and 2D aperture definitions.
All the YSOs in our sample were observed by nodding the targets along the slit in patterns made up of AB or BA pairs.
An A0V telluric standard star was observed at a similar airmass within a period of two hours before or after the science target. 
Flexure between the target and A0V observations was sub-pixel \citep{mace16}.
The airmass during the IGRINS observations ranged from 1.00 to 2.17. We used the IGRINS exposure time calculator\footnote{\url{https://wikis.utexas.edu/display/IGRINS/SNR+Estimates+and+Guidelines}} to estimate the exposure time of each object. The total exposure time varied from minutes to hours to achieve a signal-to-noise ratio (SNR)$>$100.

All IGRINS data were reduced using the IGRINS pipeline \citep*{lee17}\footnote{\url{https://github.com/igrins/plp/tree/v2.1-alpha.3}} which produces a telluric
corrected spectrum with the wavelength solution derived from OH night sky emission lines
at shorter wavelengths and telluric absorption lines at wavelengths greater than 2.2~$\mu$m. 
Telluric correction was performed by dividing the target spectrum by an A0V spectral standard and multiplying by a standard Vega model. 
We corrected for any sub-pixel shifts between the target and A0V observations (due to instrument flexure) by aligning the spectra at a strong telluric absorption feature. 
Finally, the wavelength solution of the target was corrected for the barycenter velocity derived using {\sc zbarycorr} \citep{wright2014} along with the Julian date at the midpoint of observation and the telescope site. The telluric correction uncertainties were propagated into the telluric-corrected spectra by combining the observed uncertainties of the target and standard spectra in quadrature.

\begin{figure*}
\includegraphics[width=\textwidth]{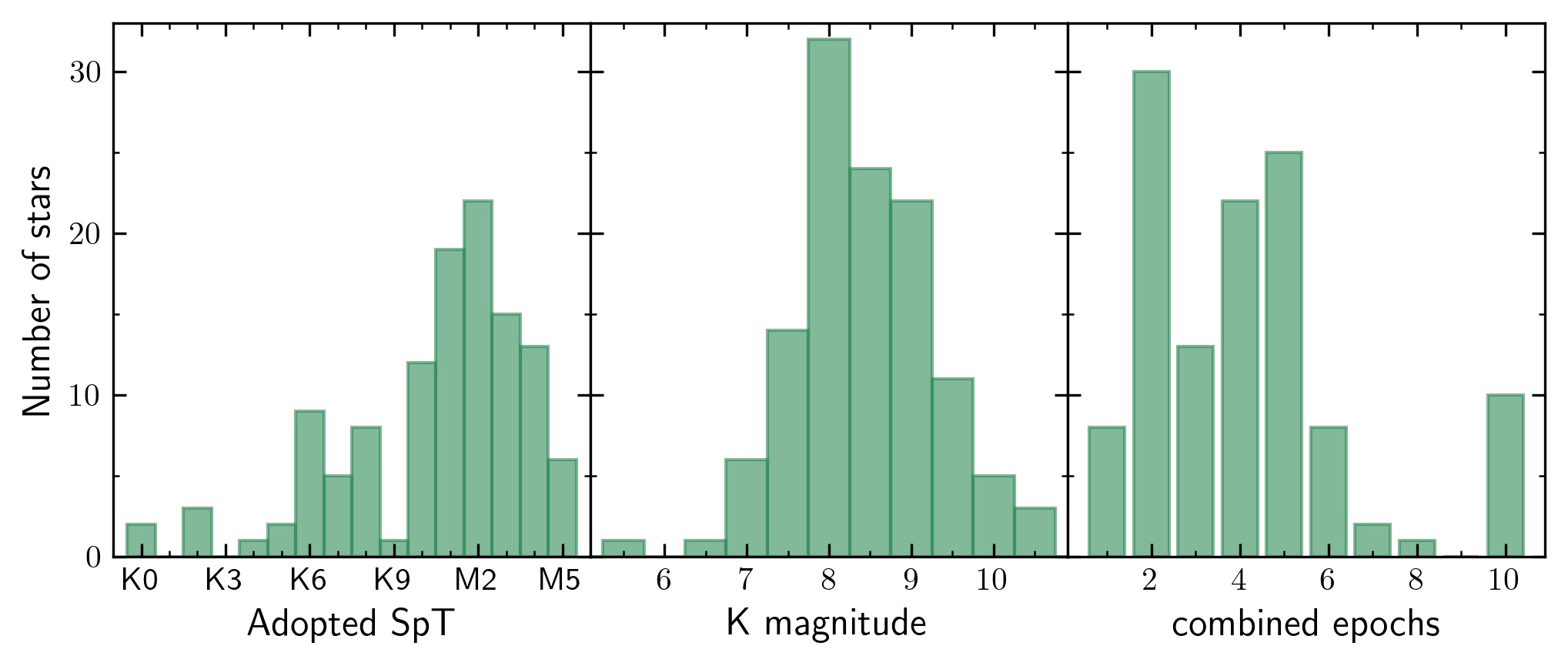}
\caption{\citet{luhman17} spectral types (left panel), 2MASS K-band magnitudes (middle panel), and the total number of combined epochs (right panel) for our sample of 119 YSOs. More than 70\% of the stars in our sample have spectral type K9 or later. Our sample also has a mean K magnitude of 8.4$\pm$0.9~mag and a median value of 4 combined epochs per target.\label{fig:histo}}
\end{figure*}

To construct the YSO sample used in this work, we looked into the reduced science spectral archive 
of IGRINS for targets classified as Taurus members by \citet{luhman2010}, 
with a spectral type (SpT) between K0 and M5 \citep{luhman17}, and a minimum average SNR of 50 in the K-band. 
The latter condition resulted in more than 500 single visit observations of 139 Taurus YSOs observed with IGRINS on the McDonald 2.7~m telescope and LDT between 2014 and 2017. 
This sample contains many of the canonical Taurus-Auriga members with K-band magnitudes $<$10~mag and according to the classifications made by \cite{kenyon95}, \cite{luhman2010}, \cite{rebull2010}, \cite{esplin2014}, and \cite{kraus17}, there are 9 Class {\sc i}, 91 Class {\sc ii} and 39 Class~{\sc iii} YSOs. 
{Finally, we visually checked the K-band spectrum of these 139 YSOs, and we excluded objects with shallow ($<$~2\%) lines due to high veiling or \vsini\ values. After this cleaning step, our final sample (see Table~\ref{tab:res}), contained 119 YSOs (84 Class {\sc ii} and 35 Class~{\sc iii}).}
Figure~\ref{fig:histo} depicts the spectral type \citep{luhman17}, 2MASS K-band magnitude \citep{cutri03}, and the number-of-epoch distributions of the 119 YSOs that define our sample.

By compiling the survey sample from the IGRINS spectral archive, we identify some biases in the sample selection. 
First, for the same effective temperature (\teff~$\lesssim$~4000~K), younger stars are brighter because of their larger radii \citep{hayashi61}. These young and bright stars were easier to observe at high spectral resolution. 
The full census of Taurus YSOs is $>$400 members \citep{luhman2018}. Our survey includes only $\sim$45\% (116 out of 258) of the Taurus YSOs between spectral types K0 and M5, since we have not observed the faintest or more widely distributed members \citep{kraus17}.
Lastly, binary stars are the specific focus of some IGRINS YSO observing programs. The combined flux of the two binary components makes them brighter, which results in higher SNRs, but their analysis is also complicated by line blending in the composite spectrum.
In this work we have identified known binaries (or multiples) in the literature and have compared the single and multiple samples. A more rigorous identification, analysis, and characterization of binarity will be the topic of a future study.

The final combined spectrum of each target was made up of between 1 and 10
epochs, with a median value of 4 epochs. If more than ten epochs were available, we chose the best ten based on a combination of airmass, SNR, and the quality of the telluric correction. 
We shifted each epoch to the wavelength rest-frame by determining the radial velocity shift of Na lines at 2.206 and 2.209~$\mu$m. Then, we produced the combined spectrum by taking a weighted average of all epochs, where the weight corresponded to the SNR at each data point. The standard deviation of the mean gave the final uncertainties per data point. 

In 2018, we observed members of the TW Hydrae Association (TWA; \citealp{katsner97}) while IGRINS visited the 8.1~m Gemini South telescope \citep{mace18}.
We reduced the TWA objects in the same fashion as the Taurus sample, but all TWA observations were single epoch and final combined spectra were not produced. 
The collection of IGRINS spectra is available from the Harvard Dataverse \citep{lopez-tau21,lopez-twa21}\footnote{\url{https://dataverse.harvard.edu/dataverse/igrins_ysos}}.

\section{Stellar parameter determination}
To determine \teff, \logg, B, and \vsini\ we used a Markov chain Monte Carlo (MCMC) algorithm comparing observations with synthetic spectra, which we describe here in detail.  

\subsection{Theoretical grid}\label{sect:grid}
We computed a four-dimensional (\teff, \logg, B, \vsini) grid of synthetic spectra using the {\sc moogstokes} code 
\citep{deen13}. {\sc moogstokes} is a customization of the one-dimensional LTE radiative transfer code {\sc moog} 
\citep{moog} that synthesizes the emergent spectrum of a star taking into account the Zeeman splitting
produced by the presence of a photospheric magnetic field. It assumes a uniform and purely radial magnetic field and uniform \teff\ and \logg, producing a disk-averaged spectrum {broadened to the user-specified spectral resolution and \vsini\ values}.

To synthesize a spectrum with {\sc moogstokes}, the program required a model atmosphere and an atomic/~molecular line list. The \teff, \logg, and metallicity of the resulting spectrum are defined by the model atmosphere, while B, \vsini, and spectral resolution are user-selected quantities.
In this work, we used the MARCS atmospheric models \citep{marcs} with solar metallicity (suitable for YSOs; \citealp{padgett96}; 
\citealp{santos08}; \citealp{dorazi11}) and the Vienna Atomic Line Database (VALD3; \citealp{vald3}). {Often, the atomic data coming from databases are not accurate enough to reproduce the Solar spectrum at high resolution, requiring modifications to some of the absoprtion line oscillator strengths and van der Wals constants. 
In this work we used the astrophysical-inferred modifications to the VALD atomic data presented by \cite{flores19}}. 
Low-mass stars have micro-turbulence values between 0 and 2~km~s$^{-1}$ \citep[e.g., ][]{gray05,reid05,bean06}, thus for this study we used a micro-turbulence of 1~km~s$^{-1}$. 
{Finally, we selected B values up to 4~kG, \vsini\ between 2 and 50~km~s$^{-1}$, and matched the IGRINS spectral resolution (R$\sim$45,000).}

Our grid of synthetic spectra covers the parameter space as follows: from 3000 to 5000~K in \teff\ 
(steps of 100~K up to 4000~K, and 250~K above 4000 K), from 3.0 to 5.0~dex in 
\logg\ (steps of 0.5~dex), from 0 to 4 kG in B (steps of 0.5~kG), and  values from 2 to 
50~km~s$^{-1}$ in \vsini\ (steps of 2~km~s$^{-1}$). {The grid steps\footnote{The grid steps of \teff\ and \logg\ are defined by the model atmosphere, while we selected the steps of B and \vsini.}  are well suited to our study as they are enough to see the effects of these parameters on the spectra. Smaller step sizes in the model grid would drastically increase the computation time without significantly altering the synthetic spectra. Instead, the grid was linearly interpolated at values between grid points. The grid of synthetic spectra does not cover the entire IGRINS spectrum, but only the spectral regions described in Section~\ref{sect:regs}. The grid of synthetic spectra is also available from Harvard Dataverse \citep{lopez-grid21}. 
} 

\subsection{Spectral Regions}\label{sect:regs}
{We employed four spectral regions in the K-band to determine the stellar parameters in this work. These regions are in four different K-band IGRINS orders. Each region is sensitive to changes in different stellar parameters, as shown by 
their use previously in K and M stars \citep[e.g.,][]{rajpurohit18,sokal18,flores19}.}

\cite{sokal18} used three spectral intervals in the K-band around Na ($\sim$2.210~$\mu$m), Ti  
($\sim$2.222~$\mu$m), and CO ($\sim$2.295~$\mu$m) features to determine \teff, \logg, and B of the star TW~Hydra. 
{Recently, \cite{flores19} included the Ca~{\sc i} lines around 2.264~$\mu$m in the determination of the stellar parameters of the young stars BP~Tau and V347 Aur with iSHELL \citep{rayner16} infrared spectra at a similar spectral resolution to IGRINS.}

All spectral lines depend, to some extent, on each of the stellar parameters. Isolating the contribution to the line profile of each parameter is only as accurate as the models and line lists. To reduce degeneracies, it was important to use a collection of spectral lines that are primarily sensitive to variations of just one stellar parameter.  

\begin{figure*}
\centering
\includegraphics[scale=0.16]{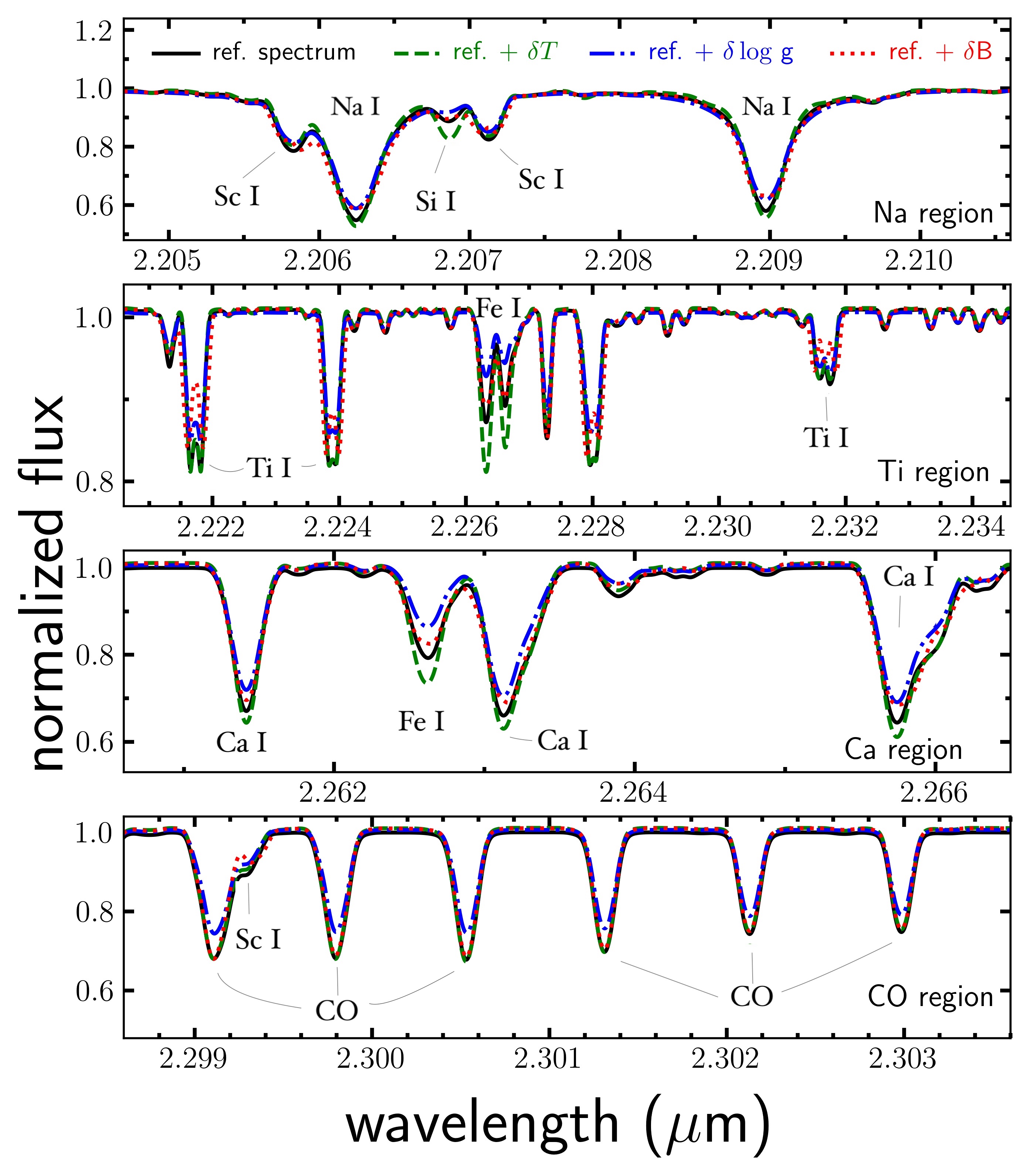}
\caption{ Dependence on stellar parameters for the selected spectral regions. We used a synthetic spectrum with \teff~=~3600~K, \logg~=~4.0~dex, B~=~2.0~kG, and \vsini~=~10~km s$^{-1}$ (solid black line) as our reference. We show the difference caused by a change of 300~K in \teff\ (green dashed line), of 0.5~dex in \logg\ (blue dash-dotted line), or 1.0~kG in B (red dotted line). In each comparison, we fixed two of the three parameters to the reference value and varied the third to a higher value. Varying parameters to lower values results in similar behavior in the opposite directions. A similar comparison can be seen in \citet{sokal18}.}
\label{fig:depen}
\end{figure*}

To assess the sensitivity of the selected spectral regions to the stellar parameters, we took the difference between two synthetic spectra varied by 300~K in \teff, 0.5~dex in \logg\ or 
1.0~kG in B. Figure~\ref{fig:depen} depicts the dependency of each spectral region on the stellar parameters. We label the four intervals as follows: 

{\it Na region} ({2.2045--2.2105}~$\mu$m): The Na doublet ($\lambda \sim$ {2.2062} and {2.2090}~$\mu$m) is the dominant feature of this region, but two Sc~{\sc i} lines ($\sim${2.2058} and {2.2071}~$\mu$m) and a 
Si~{\sc i} line ($\sim${2.2068}~$\mu$m) are also present. The Na lines show sensitivity to the magnetic field, while the Sc~{\sc i} and Si~{\sc i} are more sensitive to \teff\ and respond in opposite ways as \teff\ increases.

{\it Ti region} ({2.2205--2.2346} $\mu$m): This region has three Ti~{\sc i} lines at $\sim${2.2217}, $\sim${2.2239}~$\mu$m, and $\sim${2.2315} that are mainly sensitive to B. There are also two Fe~{\sc i} lines mainly sensitive to changes in \teff. The Ti lines present greater Zeeman splitting when the B-field increases. 

{\it Ca region} ({2.2606--2.2664}~$\mu$m): This region includes three Ca~{\sc i} lines at $\sim${2.2614, 2.2631, 
and 2.2657}~$\mu$m and one Fe~{\sc i} at $\sim${2.2626}~$\mu$m. These lines together help us to determine \teff\ and \logg\ because their dependency on B is small.

{\it CO  region} ({2.2986--2.3036}~$\mu$m): This region contains six CO lines sensitive to \logg, and a weak Sc~{\sc i} line ($\sim${2.2993}~$\mu$m) sensitive to B. As the \logg\ increases the CO line depths decrease. This CO region is smaller than others have used, which provides better continuum flattening.

The combined use of the four regions located in the K-band allows for the accurate simultaneous determination of \teff, \logg, B, and \vsini\ of our YSOs sample.

\subsection{Determining Stellar Parameters}\label{sect:detpars}
To determine stellar parameters we continuum normalized each spectral region through an interactive python script. First, the wavelength range of the spectral region is equally divided into n bins. Then, we computed the median ($\mu_{\rm flux}$) and the standard deviation ($\sigma_{\rm flux}$) of the flux in each bin; we then excluded points out of $\mu_{\rm flux}\pm0.5\sigma_{\rm flux}$, and we recomputed the median flux of the bin ($\mu_{\rm clip}$). Finally, we fit a polynomial of order k to the $\mu_{\rm clip}$ of the n bins. The values of n and k vary between objects and spectral regions and could be interactively modified to obtain a better normalization. We typically used between 10 and 20 bins and between 1 and 4 for the polynomial order. 

Then, we carried out a MCMC analysis, as implemented in the code {\it emcee} \citep{emcee}, using the four K-band spectral regions (Na, Ti, Ca, and CO). We compared observed and synthetic spectra by allowing \teff, \logg, B, \vsini, and K-band veiling ($r_{\rm K}$) to vary along with small continuum ($<$~3\%) and wavelength ($<$~0.6\AA) offsets.
In each MCMC trial we linearly interpolated within the four-dimensional (\teff, \logg, B, \vsini) synthetic spectral grid described in Section~\ref{sect:grid}, to obtain the corresponding spectrum with the sampled set of parameters. The interpolated synthetic spectrum was then artificially veiled and re-normalized for each region following this equation:
\begin{equation}\label{eq:for_veil}
    F_v = \frac{F + r_{\rm K}}{1+r_{\rm K}}
\end{equation}
where $F_v$, $F$, and $r_{\rm K}$ are the veiled flux, the synthetic flux, and the K-band veiling value \citep{basri90}. A single, best-fit veiling value was determined for all the K-band wavelength regions at once. Note that the $r_{\rm K}$ values determined with our MCMC analysis are not strictly a veiling value, but a combination of the true veiling with secondary systematic offsets between the model and target continua.

We then evaluated each MCMC with a likelihood function based on the sum of the $\chi^2$ statistics divided by the number of pixels in each region. Finally, from the posterior probability distributions of the MCMC, we took the 50th percentile as the most likely value for the stellar parameters. 

\begin{figure*}
    \centering
    \includegraphics[width=\textwidth]{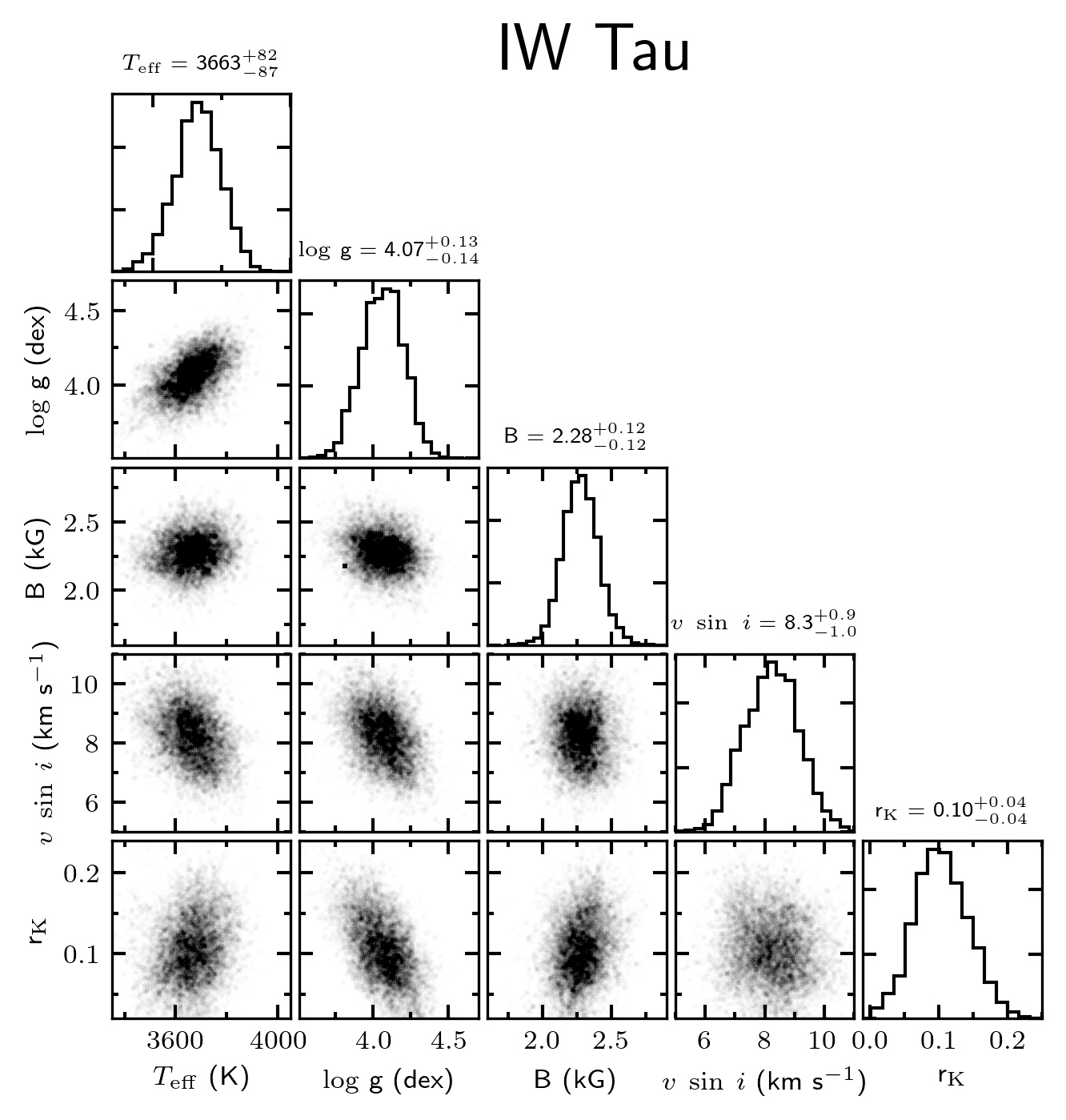}
    \caption{Posterior distributions and degeneracies (after removing the burn-in phase) of \teff, \logg, B, \vsini, and r$_{\rm K}$ determined for IW Tau. The median value and the 1$\sigma$ uncertainties are printed above each distribution. \label{fig:corner}}
\end{figure*}

In Figure~\ref{fig:corner}, we show the posterior distributions of the MCMC analysis for the star IW~Tau as an illustrative example. { Figure~\ref{fig:corner} also shows the dependence between stellar parameters like \logg\ and r$_{\rm K}$. 
A good fit to the observation can be achieved with a slightly lower \logg\ , or a higher r$_{\rm K}$ because both \logg\ or veiling varies the CO line depths.
The degeneracy between these parameters adds to the uncertainty in each quantity separately.}

\subsection{Uncertainties}\label{sect:err}
The 16th and 84th percentiles of the posterior probability distributions of the stellar
parameters represent the formal uncertainties of our fitting method. 
Most of the uncertainties were found to be nearly symmetric, and we report the larger of the two percentiles as the corresponding symmetric uncertainty ($\sigma_{\rm fit}$) for each parameter. The median fit uncertainties are 152~K in \teff, 0.25~dex in \logg, 0.54~kG in B, 1.9~km~s$^{-1}$ in \vsini, and 0.10 in r$_{\rm K}$.

{Additionally, we quantified the systematic uncertainties ($\sigma_{\rm sys}$) by comparing our stellar parameters to previously published values.} Stars with precise physical parameters provided by interferometric observations help mitigate model dependency by calibrating relationships between \teff\ and stellar radius. The works of \cite{mann13,mann15} and \cite{newton15} used 20$+$ stars with interferometric measurements to calibrate their model-independent relationships with $\sim$150~K precision. {To quantify the $\sigma_{\rm sys}$ for \teff\ and \logg, we determined the stellar parameters of seven field M stars with metallicity close to the solar value ([Fe/H]=0.0$\pm$0.10~dex) from \cite{mann13}.

We used the same fitting routines as for the YSO sample, but assuming r$_{\rm K}$~=~0.0 since these stars are on the main sequence and diskless.} In Table~\ref{tab:calib}, we compared our \teff\ and \logg\ values with those obtained by \cite{mann13}.
They determined \teff, [Fe/H], mass (M) and radius (R), which we used in the following equation to compute their \logg\ values:
\begin{equation}
\log {\rm g}_{\rm cal} = \log (\rm M/\rm M_\odot) - 2\log (\rm R / \rm R_\odot) + 4.437 
\end{equation}

\begin{deluxetable*}{llcrrrcccrcrrrcc}
\tabletypesize{\scriptsize}   
\rotate
\tablewidth{0pt}
 \tablecaption{Compiled information for the 119 stars in our Taurus YSO sample, including \teff, \logg, B, and \vsini. Columns 1-8 provide target information from the literature, while Columns 9-16 present information and parameters from this work. {The Ref column identifies the source of the class and binary (objects with companions closer than 2 arcseconds) classification. The SpT and the K magnitude comes from \citet{luhman17} and \citet{cutri03}, respectively.} Uncertainties are the 1$\sigma$ total (fit + systematic) uncertainties.  The N column is the number of epochs combined for each object. The SNR$_{\rm K}$ column is the median SNR of the combined spectrum for the spectral regions used in this work. {The flag column equals 0, 1, 2 or 3 indicates good, acceptable,  outside/edge of grid, or poor determinations.}The full version of this table is available in the online version of the paper.\label{tab:res}}
\tablecolumns{16}
\tablehead{
\colhead{2MASS ID} & \colhead{Alt. Name} & \colhead{K} & \colhead{SpT} & \colhead{Class} & \colhead{Ref} & \colhead{Binary} & \colhead{Ref} & \colhead{N}&\colhead{SNR$_{\rm K}$}&\colhead{flag}& \colhead{\teff} & \colhead{\logg} & \colhead{B} & \colhead{\vsini} & \colhead{$r_{\rm K}$}  \\
\colhead{}& \colhead{} &\colhead{(mag)}& \colhead{} & \colhead{} & \colhead{} & \colhead{}& \colhead{}& \colhead{}& \colhead{} & \colhead{}& \colhead{(K)}& \colhead{(dex)}& \colhead{(kG)}& \colhead{(km s$^{-1}$)}&\colhead{}}
\startdata
J04141760+2806096 & [BCG93] 1 & 9.9 & M5.0 & II & 1,2,3,4 & ? & -- & 2 & 172 & 0 & 3281$\pm$197 &  3.88$\pm$0.40 & $<$1.09 & 9.9$\pm$2.9 & 1.10$\pm$0.21 \\  
J05075496+2500156 & [BCG93] 12 & 10.4 & M3.7 & II & 2,4 & N & 16 & 2  & 64 & 3 & $--$ & $--$ & $--$ & $--$ & $--$ \\ 
J04182909+2826191 & [SS94] V410 Anon 25 & 9.9 & M3.5 & III & 2,4,5 & ? & -- & 1 & 87 & 1 & 3619$\pm$328 &  4.05$\pm$0.59 & $<$2.32 & 30.7$\pm$7.4 & 0.39$\pm$0.20 \\  
J04174965+2829362 & [SS94] V410 X-ray 1 & 9.1 & M3.7 & II & 2,3,4 & N & 16 & 2 & 182 & 0 & 3315$\pm$129 &  3.37$\pm$0.25 & 1.22$\pm$0.33 & 9.7$\pm$2.0 & 0.28$\pm$0.06 \\  
J04183444+2830302 & [SS94] V410 X-ray 2 & 9.2 & M0.0 & II & 2,4 & ? & -- & 2 & 182 & 0 & 4231$\pm$143 &  4.53$\pm$0.27 & 1.84$\pm$0.49 & 10.9$\pm$2.9 & 0.44$\pm$0.08 \\  
J04345542+2428531 & AA Tau & 8.0 & M0.6 & II & 1,2,3,4 & N & 16 & 10 & 356 & 0 & 3751$\pm$171 &  3.87$\pm$0.26 & 2.16$\pm$0.41 & 12.5$\pm$2.3 & 1.56$\pm$0.12 \\  
J04191583+2906269 & BP Tau & 7.7 & M0.5 & II & 1,2,3,4 & N & 16 & 7 & 353 & 0 & 3719$\pm$131 &  4.01$\pm$0.25 & 2.19$\pm$0.37 & 9.9$\pm$2.3 & 1.36$\pm$0.11 \\  
J04335200+2250301 & CI Tau & 7.8 & K5.5 & II & 2,3,4 & N & 16 & 10 & 931 & 0 & 3951$\pm$94 &  3.77$\pm$0.17 & 1.95$\pm$0.31 & 12.5$\pm$1.9 & 2.34$\pm$0.09 \\  
J04265440+2606510 & CoKu FV Tau c & 8.9 & M2.5 & II & 1,2,3,4 & Y & 24 & 6 & 302 & 0 & 3456$\pm$136 &  3.53$\pm$0.23 & $<$2.01 & 38.6$\pm$2.4 & 0.33$\pm$0.06 \\  
J04355349+2254089 & CoKu HP Tau G3 & 8.8 & M0.6 & III & 1,2,3,4,5 & Y & 16 & 6 & 239 & 0 & 3760$\pm$164 &  3.84$\pm$0.26 & $<$2.65 & 43.3$\pm$2.9 & 0.25$\pm$0.07 \\  
J04354093+2411087 & CoKu Tau 3 & 8.4 & M0.5 & II & 1,2,3,4 & N & 16 & 5 & 249 & 0 & 3665$\pm$154 &  3.85$\pm$0.26 & 2.33$\pm$0.34 & 8.2$\pm$2.1 & 0.77$\pm$0.08 \\  
J04411681+2840000 & CoKu Tau 4 & 8.7 & M1.1 & II & 1,2,4 & Y & 13 & 6 & 254 & 0 & 3706$\pm$121 &  3.70$\pm$0.21 & $<$1.90 & 26.3$\pm$2.2 & 0.22$\pm$0.04 \\  
J04144786+2648110 & CX Tau & 8.8 & M2.5 & II & 1,2,3,4 & N & 16 & 2 & 152 & 0 & 3520$\pm$180 &  3.61$\pm$0.30 & $<$1.39 & 20.5$\pm$2.5 & 0.28$\pm$0.08 \\  
J04173372+2820468 & CY Tau & 8.6 & M2.3 & II & 1,2,3,4 & N & 16 & 2 & 203 & 0 & 3445$\pm$155 &  3.73$\pm$0.28 & $<$1.31 & 10.8$\pm$2.3 & 0.81$\pm$0.13 \\  
J04183158+2816585 & CZ Tau & 9.4 & M4.2 & II & 1,2,3,4 & Y & 18 & 4 & 220 & 1 & 3385$\pm$211 &  3.83$\pm$0.35 & $<$1.70 & 31.4$\pm$3.1 & 0.64$\pm$0.12 \\  
J04183112+2816290 & DD Tau & 7.9 & M3.5 & II & 1,2,3,4 & Y & 6 & 4 & 251 & 0 & 3250$\pm$207 &  3.47$\pm$0.37 & $<$1.48 & 16.9$\pm$3.1 & 1.82$\pm$0.26 \\  
J04215563+2755060 & DE Tau & 7.8 & M2.3 & II & 1,2,3,4 & N & 16 & 4 & 251 & 0 & 3463$\pm$158 &  3.45$\pm$0.27 & $<$1.01 & 9.4$\pm$2.2 & 1.11$\pm$0.11 \\  
J04270280+2542223 & DF Tau & 6.7 & M2.0 & II & 1,2,3,4 & Y & 24 & 4  & 293 & 3 & $--$ & $--$ & $--$ & $--$ & $--$ \\ 
J04270469+2606163 & DG Tau & 7.0 & K7.0 & II & 1,2,3,4 & N & 16 & 5 & 553 & 2 & 3969$\pm$216 &  3.20$\pm$0.28 & $<$1.78 & 26.9$\pm$4.3 & 4.82$\pm$0.94 \\  
J04294155+2632582 & DH Tau & 8.2 & M2.3 & II & 1,2,3,4 & N & 16 & 10 & 399 & 0 & 3477$\pm$125 &  3.89$\pm$0.22 & 2.21$\pm$0.32 & 8.4$\pm$2.1 & 1.24$\pm$0.10 
\enddata
\tablereferences{(1)\citet{kenyon95};
(2)\citet{luhman2010};
(3)\citet{rebull2010};
(4)\citet{esplin2014};
(5)\citet{kraus17};
(6)\citet{bouvier92}; 
(7)\citet{correia06};
(8)\citet{duchene99}; 
(9)\citet{dyck82}; 
(10)\citet{ghez93};
(11)\citet{haas90}; 
(12)\citet{herbig88}; 
(13)\citet{ireland08}; 
(14)\citet{konopacky07}; 
(15)\citet{kraus06};
(16)\citet{kraus11}; 
(17)\citet{leinert91}; 
(18)\citet{leinert93}; 
(19)\citet{mathieu91};
(20)\citet{mathieu96};
(21)\citet{richichi99}; 
(22)\citet{ruiz16};
(23)\citet{schaefer14};
(24)\citet{simon92};
(25)\citet{walter88};
(26)\citet{weintraub89}.}
\end{deluxetable*}

{ We found an average difference in \logg\ (ours - Mann) of $-$0.04~dex and a standard deviation of the residuals of 0.13~dex. For \teff\ we found a mean difference of -69~K and a standard deviation of the residuals of 75~K}. Therefore, we assigned 0.13~dex and 75~K as the systematic uncertainties for \logg\ and \teff, respectively.

\begin{figure}
    \centering
    \includegraphics[width=\columnwidth]{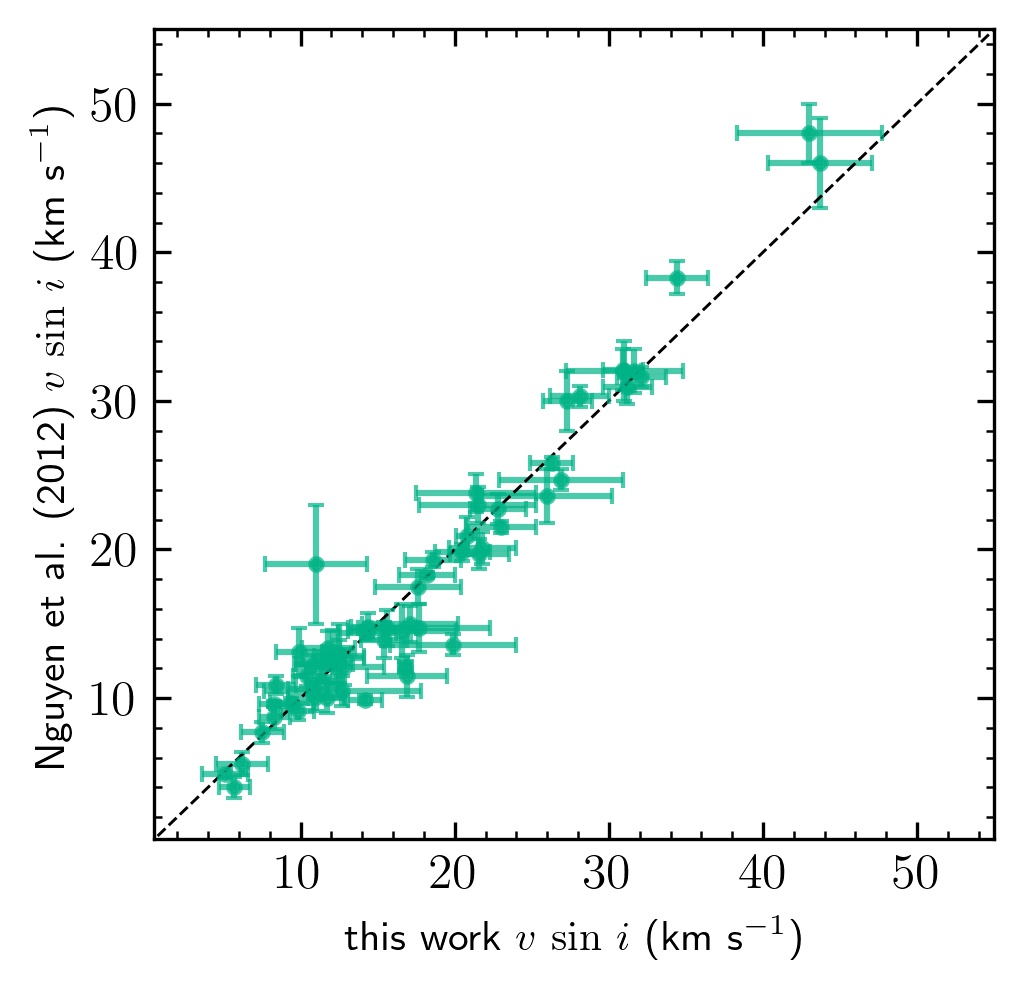}
    \caption{Comparison of \vsini\ values between \citet{nguyen12} and this work for 60 YSOs in common. The dashed line shows the one-to-one relation. The mean difference between both studies is 0.08$\pm$0.3~km~s$^{-1}$, and we found a correlation coefficient of 0.97. Our error bars are the fit uncertainties. \label{fig:c_vrot}}
    
\end{figure}

{To obtain the systematic uncertainty on \vsini, we compared measurements for 52 YSOs in common with \cite{nofi21}. In that work, they determined the \vsini\ of pre-main-sequence stars using the same single-epoch, uncombined IGRINS K-band spectra that we have used. Their method used the cross-correlation technique outlined in \cite{hartmann86} and \cite{soderblom89} along with synthetic spectra. 
In brief, a stellar spectrum with rotationally broadened lines was cross-correlated against an unbroadened spectrum. Then the width of the cross-correlation function is used to measure the rotational broadening of the spectrum. 
Generally, our \vsini\ values were in good agreement with those of \cite{nofi21}.
We found a mean difference between the \vsini\ values (ours - Nofi) of 0.2~km~s$^{-1}$ and a standard deviation of the residuals of 1.7~km~s$^{-1}$. We assigned this 1.7~km~s$^{-1}$ as the \vsini\ systematic uncertainty.

{Over 85\% of our \vsini\ values are $\leq$30~km~s$^{-1}$, with the median value being 16.6~km~s$^{-1}$}. In Figure~\ref{fig:c_vrot} we compare \vsini\ measurements for 60 YSOs in common with \cite{nguyen12}. They used high-resolution (R=60,000) optical (4800--9300~\AA) spectra to 
determine \vsini\ for pre-main-sequence stars in the Chamaeleon~{\sc i} and Taurus-Auriga star-forming regions.
Their method compared each echelle order of the target spectrum with slow-rotating template stars, which were artificially broadened to different values of \vsini\ to find the best fit.  Finally, they removed 2.7$\sigma$ outliers  around  the  mean  value  and computed a weighted  average  over  all  the  remaining  echelle  orders \citep[for more details see also][]{nguyen09}. We found our \vsini\ values in agreement with those of \cite{nguyen12}, with an mean difference of 0.08~km~s$^{-1}$, and a error on the mean of 0.3~km~s$^{-1}$.}

{ Finally, we collected 12 determinations (for 11 YSOs) of B-field strength from the literature \citep{johns-krull07,sokal20,flores20} to assess the systematic uncertainty of our B values (see Table~\ref{tab:mag}). We found a mean difference in the residuals (ours - literature) of $-0.27$~kG and a standard deviation of 0.26~kG. We assigned 0.26~kG as our systematic uncertainty in the B-field measurements.}

The systematic uncertainties were added in quadrature to the fit errors to compute the total error of each parameter ($\sigma_{\rm tot}$), which we report in Table~\ref{tab:res}.

\begin{deluxetable}{lcccc}
\tablewidth{0pt}
 \tablecaption{Effective temperature and surface gravity determined for our calibration sample as well as those determinations of \citet{mann13}. The fit errors are reported on our \teff\ and \logg\ determinations.\label{tab:calib}}
\tablehead{ \colhead{} & \multicolumn{2}{c}{\citet{mann13}} & \multicolumn{2}{c}{this work} \\
\colhead{Star} & \colhead{\teff} & \colhead{\logg} & \colhead{\teff} & \colhead{\logg} \\
\colhead{} & \colhead{(K)} & \colhead{(dex)} & \colhead{(K)} & \colhead{(dex)} }  

\startdata 
GJ 338A&3953$\pm$41&4.71$\pm$0.05 & 3813$\pm$56 & 4.68$\pm$0.09 \\
GJ 338B&3926$\pm$37&4.72$\pm$0.04 & 3785$\pm$54  & 4.69$\pm$0.08 \\
GJ 436 &3520$\pm$66&4.77$\pm$0.06 & 3587$\pm$59  & 4.90$\pm$0.11 \\
GJ 570A&4588$\pm$58&4.57$\pm$0.07 & 4536$\pm$50  & 4.36$\pm$0.11 \\
GJ 702B&4475$\pm$33&4.66$\pm$0.04 & 4391$\pm$38  & 4.50$\pm$0.14 \\
GJ 809 &3744$\pm$27&4.72$\pm$0.04 & 3604$\pm$63  & 4.59$\pm$0.09 \\
GJ 887 &3695$\pm$35&4.78$\pm$0.05 & 3699$\pm$67  & 4.95$\pm$0.08\\
\enddata
\end{deluxetable}

\begin{deluxetable}{lccc}
\tablewidth{0pt}
 \tablecaption{Literature and this work magnetic field strength determinations for 11 YSOs. The reference column are JF07~=~\citet{johns-krull07}, F19~=~\citet{flores19}, and S20~=~\citet{sokal20}. We report the fit uncertainties on our B values.\label{tab:mag}}
  
\tablehead{ \colhead{Star} & \colhead{B literature} & \colhead{Ref.} & \colhead{B this work} \\
\colhead{} & \colhead{(kG)} & \colhead{} & \colhead{(kG)} }
\startdata
AA Tau &2.78$\pm$0.28      &JK07      &2.16$\pm$0.32\\ 
BP Tau &2.17$\pm$0.22      &JK07      &2.19$\pm$0.27\\ 
BP Tau &2.50$\pm$0.16      &F19       &2.19$\pm$0.27\\ 
CI Tau &2.15$\pm$0.15      &S20       &1.95$\pm$0.16\\ 
CY Tau &1.16$\pm$0.12      &JK07      &1.31$\pm$0.37\\ 
DE Tau &1.12$\pm$0.11      &JK07      &1.01$\pm$0.24\\ 
DH Tau &2.68$\pm$0.27      &JK07      &2.21$\pm$0.18\\ 
DK Tau &2.64$\pm$0.26      &JK07      &2.55$\pm$0.58\\ 
DN Tau &2.00$\pm$0.20      &JK07      &1.54$\pm$0.26\\ 
GI Tau &2.73$\pm$0.27      &JK07      &1.92$\pm$0.36\\ 
GK Tau &2.28$\pm$0.23      &JK07      &2.05$\pm$0.87\\ 
GM Aur &2.22$\pm$0.22      &JK07      &2.11$\pm$0.22\\ 
\enddata
\end{deluxetable}


\section{Results and analysis}
\begin{figure}
    \centering
    \includegraphics[width=\columnwidth]{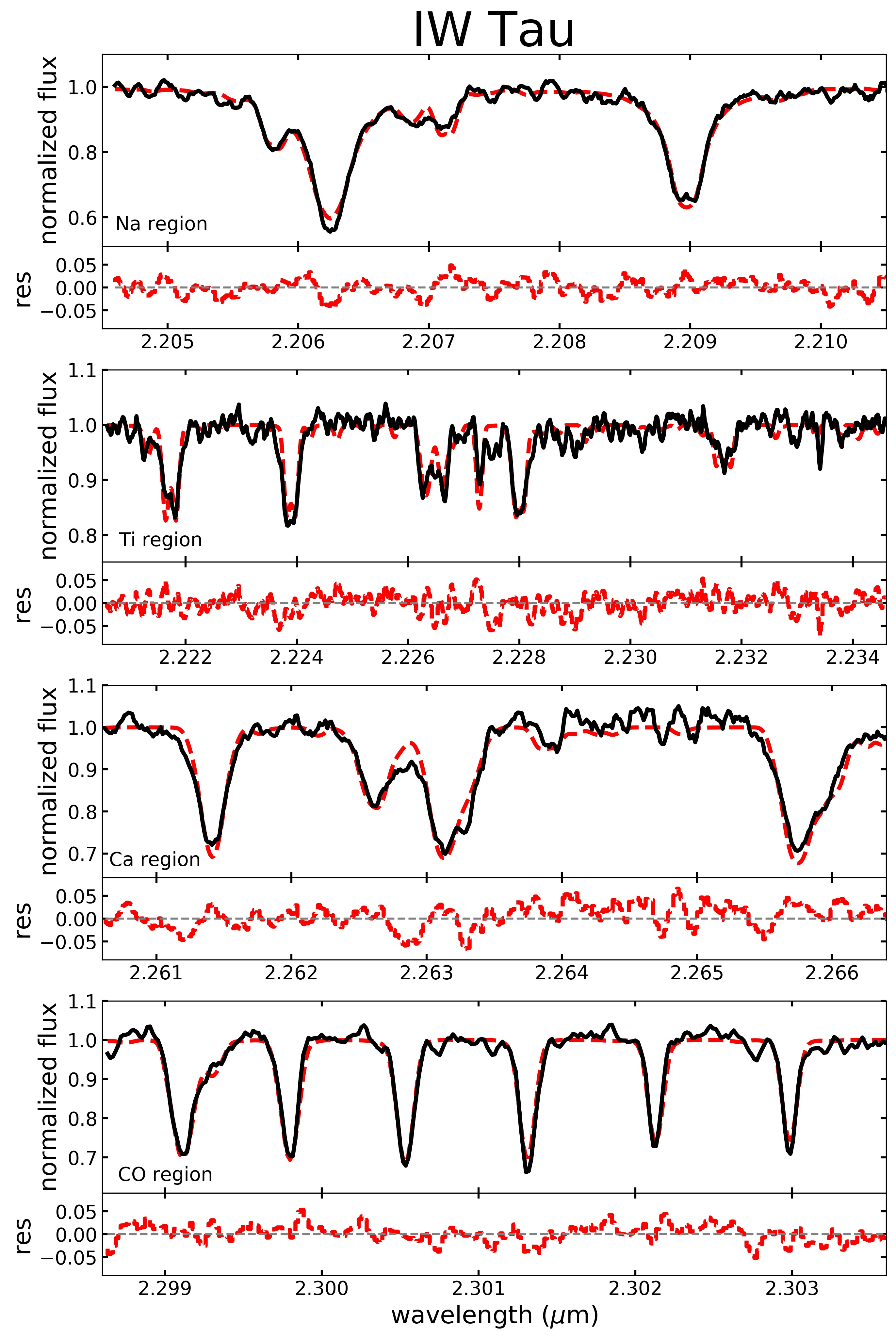}
    \caption{The four IGRINS K-band spectral regions of the star IW Tau (solid black line). We include the best-fit synthetic spectrum (red dashed line) with \teff~=~3663~K, \logg~=~4.07 dex, B~=~2.28~kG, \vsini~=~8.3~km~s$^{-1}$, and r$_{\rm K}$~=~0.10. In the bottom panels, we present the residuals between the observed and the synthetic spectra. The level of agreement between the best-fit-parameter synthetic spectrum and the IW Tau observed spectrum was considered a good determination. \label{fig:bestfit}}
\end{figure}

We plotted the spectral regions used in our analysis for all the observed and the best-fit spectra, as in Figure~\ref{fig:bestfit}, to categorize the quality of our parameter determinations. 
Through a ranking of the $\chi^2$ statistics and visual inspection, we marked each set of parameters in Table~\ref{tab:res} with a numerical {flag equal to 0, 1, or 3 if they are good, acceptable, or poor determinations, respectively. We also computed the lower and upper limits of each parameter using its total uncertainty, to look for values outside or at the edge of our grid. We find that six lower limits of \logg\ meet this criterion. We flagged these parameters with a flag equal to 2 and have retained these measurements in our analysis since they are otherwise classified as `good' determinations.} We found that our parameter determination method needs a minimum median SNR~$\gtrsim$~80 to produce reliable results. A median SNR~$\gtrsim$~150 is generally required for good parameter determinations.
In total, {we identified 80, 24, 9, and 6 stars whose parameters are good, acceptable, poor, and outside/edge of grid}, respectively. We exclude poor determinations in further analysis, and we will refer to the 110 remaining YSO measurements as our Taurus sample.

\subsection{Magnetic fields}\label{sect:field}
While Zeeman line splitting effects are most significant in the Ti region, the Na region is also sensitive to variations in the B field (see Figure~\ref{fig:depen}). However, our ability to measure Zeeman line separations is primarily a function of the line widths, which are dominated by \vsini.  
\cite{hussaini20} measured the line splitting of the Na and Ti spectral regions in IGRINS data and the {\sc moogstokes} spectra employed in this work. 
That work found that the retrieval of B is limited to $\sim$\vsini/10 in the Ti region and $\sim$\vsini/5 in the Na region.
In other words, an object with \vsini~=~15~km~s$^{-1}$ must have B~$>$1.5~kG for Zeeman splitting to be measurable in the Ti region and $>$3~kG to be measurable in the Na region. 
Additionally, the IGRINS spectral resolution limits the minimum measurable B field from line splitting to $\sim$0.7~kG. For this work, we conservatively considered a detection limit of B~$\geq$~(\vsini/8)~kG, and no lower than 1~kG. Values of B field lower than 1~kG can not be distinguished from a non-detection at the spectral resolution of IGRINS when simultaneously fitting for other model parameters.
We found that 41 of our B field determinations meet the detection threshold. The remaining upper-limit measurements are provided in Table~\ref{tab:res}.

\subsection{Impact of B on the stellar parameters }
\begin{figure*}
    \centering
    \includegraphics[width = \textwidth]{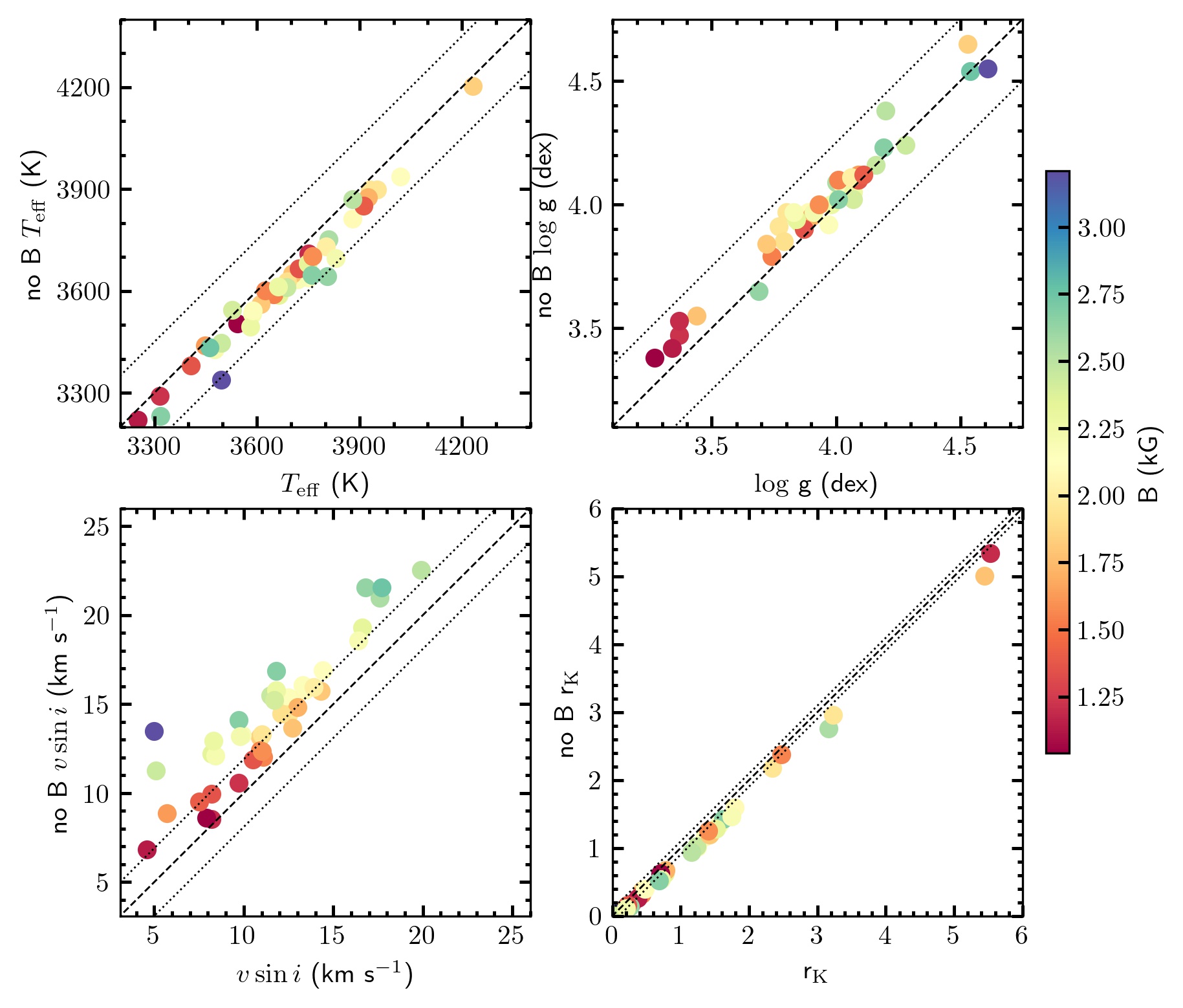}
    \caption{Stellar parameters determined with and without taking into account the effects of B-field strength. The dashed line is the one-to-one relation, while the dotted lines demarcate the median fit error intervals namely: $\pm$150~K, $\pm$0.25~dex, $\pm$1.9~km~s$^{-1}$, and $\pm$0.10 for \teff, \logg, \vsini, and $r_{\rm K}$, respectively. The data are color-coded by the determined strength of B.}
    \label{fig:comp-b}
\end{figure*}

To evaluate the impact of B on \teff, \logg, \vsini, and $r_{\rm K}$, we fixed B to 0~kG and redetermined the stellar parameters for the 41 YSOs with B-field detections in Table~\ref{tab:res}.
In Figure~\ref{fig:comp-b} we compare the stellar parameters determined in our primary analysis with those obtained by disabling the B parameter. We further consider the low B (objects with B~$\leq$~2~kG) and the high B (B~$>$~2~kG) results in this comparison. 

We found that all the \teff\ values determined without considering the B-field effects are systematically colder than those found considering the B-field. For low B (18 objects), we found a mean difference of $-$42$\pm$4~K, while for high B (23 objects), this difference increases slightly to $-$73$\pm$9~K. Both values are well within the median total (fit+systemic) \teff\ error, which is $\sim$170~K.
Similarly, r$_{\rm K}$ is lower when B is not taken into account, with a mean difference of $-$0.14$\pm$0.02 for both the low and high B bins.
Contrary to \teff\ and r$_{\rm K}$, most of the \logg\ values are higher by 0.08$\pm$0.01~dex and 0.03$\pm$0.01~dex for the low and high B bins, respectively, when not considering B-fields in the fit. These \logg\ differences are within the \logg\ median total (fit+systemic) error of 0.28~dex. 
It is likely that the degeneracy between \logg\ and $r_{\rm K}$ discussed in Section~\ref{sect:detpars} is why \logg\ is overestimated and $r_{\rm K}$ is underestimated when B is not included in the parameter determinations. 

The most prominent effect of ignoring the B-field strength when determining parameters is seen in the \vsini\ values. All \vsini\ determinations are higher without B-field included in the fit, and the difference is greater for larger values of B. We found for the low-B bin a mean difference of 1.6$\pm$0.2~km~s$^{-1}$. For the high-B bin, the difference increases to 3.8$\pm$0.3~km~s$^{-1}$. This trend is explained by excess rotational broadening of the synthetic spectral lines to try and replicate the B-field induced Zeeman line-splitting. The absence of B-field considerations by \cite{nguyen12} and \cite{nofi21} is likely responsible for some of the differences between those works and ours, which adds to the systematic uncertainties.

The previous tests suggest that it is possible to obtain suitable stellar parameters, even when excluding the B-field from the determination. 
However, the parameters will have systematic offsets, with magnetically active YSO parameters being the most impacted.


\subsection{Effective Temperature}
\begin{figure*}
    \centering
    \includegraphics[width=\textwidth]{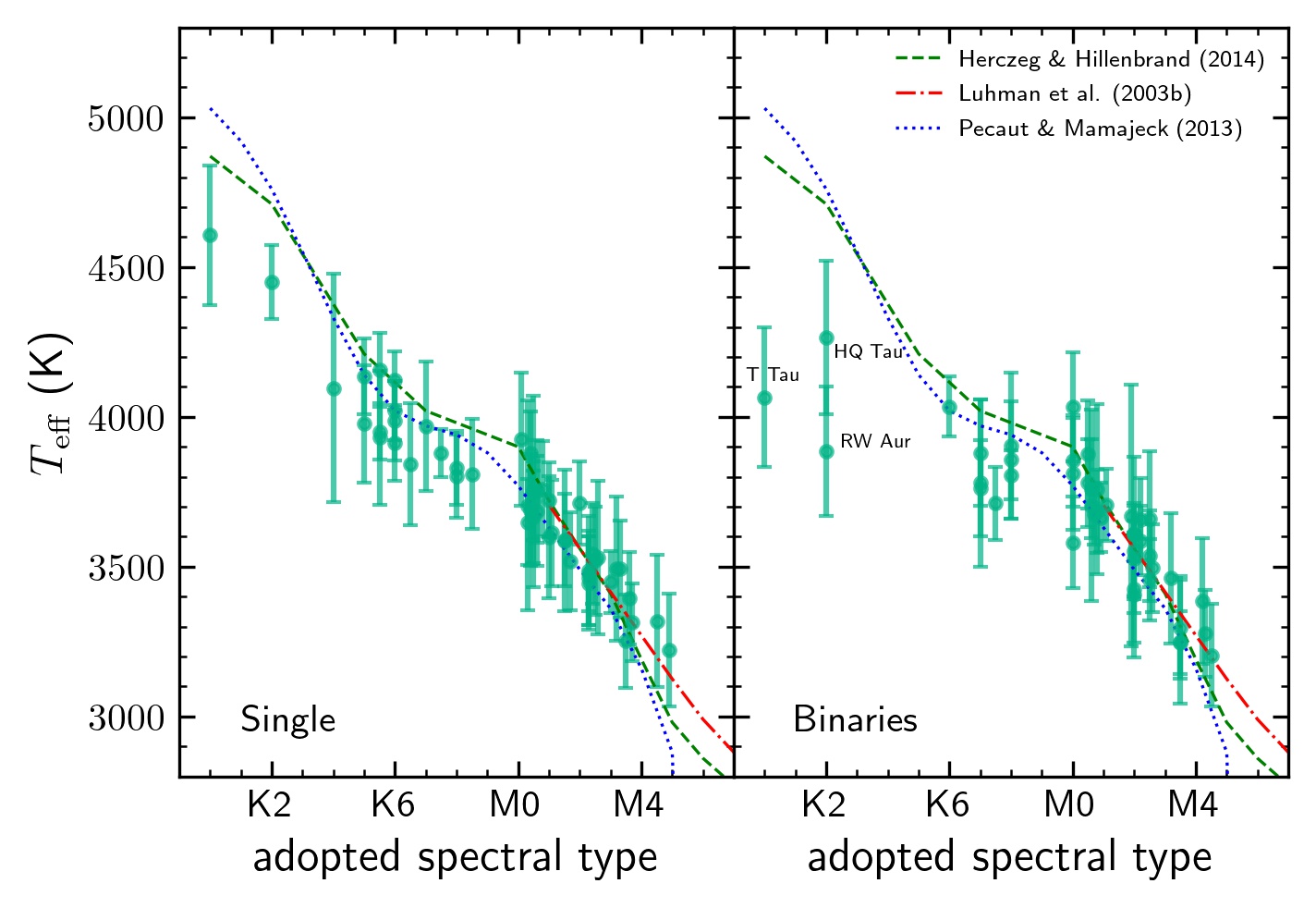}
    \caption{Effective temperature as a function of \citet{luhman17} spectral type for our Taurus-Auriga known single and binary (multiple) YSOs. The dashed, dotted-dashed, and dotted lines are the temperature scales of  \citet{herczeg14}, \citet{luhman03b} and \citet{pecaut13}, respectively. Typically the error on the spectral type is about 1 subclass. There is good agreement between our temperatures for M0--M4 single and K6--M5 binary stars with the published temperature scales. For single K stars, our \teff\ values are cooler than the published temperature scales. The binary stars HQ Tau, T Tau, and RW Aur present the largest discrepancies. Our error bars are the total (fit + systematic) \teff\ uncertainties. \label{fig:comp_tem}} 
    \end{figure*}
    
\cite{kenyon95} presented a conversion between spectral type and \teff\ for young stars, based on the work of \cite{schmidt82} and \cite{straivzys92}. Shortly after, \cite{luhman99} developed a new temperature scale for young M stars between the giant and dwarf scales. 
However, improvements made in the last two decades in the modeling of
cool star atmospheres, and the development of new instruments, have resulted in several studies that have determined effective temperatures (or spectral type) either for field (e.g., \citealp{casagrande08}; \citealp{rajpurohit13}; \citealp{rajpurohit18}; \citealp{lopezval19}) or for young (e.g., \citealp{luhman03}; 
\citealp{luhman03b}; \citealp{herczeg14}; \citealp{cottar14}; \citealp{yao18}) K \& M stars. 

In Figure~\ref{fig:comp_tem} we compare the \teff\ values we determined to spectral types from \citet{luhman17}, along with three different temperature scales of pre-main-sequence stars \citep{luhman03b, herczeg14, pecaut13}. To analyze how multiplicity plays a role in determining \teff, we divided our sample into single (51) and binary (44) stars based on classifications from the literature. These identifications are reported in Table~\ref{tab:res}. {We consider a star a binary if it has a companion closer than 2 arcseconds because the resultant IGRINS spectrum will contain some amount of flux from both components. 

The amount of contamination from a binary companion is not easily quantified.
An equal brightness binary with no velocity offsets will have composite line profiles, often producing poor fits to the data. 
If the binary has equal brightness and radial velocity offsets (a double-lined spectroscopic binary), then line blending in the combined spectrum will often result in poor parameter determinations.
Generally, equal brightness binaries have been excluded from our analysis by visual inspection in the initial sample selection and at the parameter-fit quality check.
Binaries with disproportionate fluxes will have the same issues as the equal flux binaries, but at lower levels. 
Stellar parameters obtained for the known binary stars should be used with caution since secondary contamination is possible, although the spectral fit is dominated by the primary star's contribution. The binary identification here allows us to look for systematic differences in the data.}

{In general our temperatures follow the scales of \cite{pecaut13}, \cite{herczeg14} and \cite{luhman03b} for both the single and binary samples. Early K-type binaries (T~Tau, RW~Aur, and HQ~Tau) show large discrepancies which is possibly due to wrong SpT classifications, as \citet{flores20} also found a cooler temperature for T~Tau (\teff=3976$\pm$90~K), cool spots or the binary nature of these YSOs biasing measurements to cooler temperatures. A more detailed multi-epoch and multi-wavelength study should be carried out at higher angular resolution to understand the large variation in temperatures for these early-K binaries.

A temperature scale for the single stars was computed using the mean value and the standard deviation within $\pm$0.5 spectral type sub-classes bins. This IGRINS temperature scale, reported in Table~\ref{tab:scale}, represents an alternative to previous published pre-main-sequence stars temperature scales, with the advantage of being obtained through a simultaneous determination of various atmospheric parameters and the inclusion of the magnetic field effects.}

\begin{deluxetable}{lccclc}
\tablewidth{0pt}
 \tablecaption{Temperature scales of \citet[][L03]{luhman03b}, \citet[][PM13]{pecaut13}, \citet[][HH14]{herczeg14} and that determined here for single YSOs. The last column correspond to the number of objects with which we computed the mean \teff\ value and the standard deviation within a $\pm$0.5 spectral type subclasses. If just one object was included in the spectral type bin, we report the value of \teff\ (and its error) found in our analysis. \label{tab:scale}}
\tablehead{ \colhead{SpT} & \colhead{L03} & \colhead{PM13} & \colhead{HH14} & \colhead{This work} & \colhead{N} \\
& \colhead{(K)}& \colhead{(K)} & \colhead{(K)} & \colhead{(K)} & }
\startdata 
K0  & --    &  5030  & 4870  &  4606$\pm$233  & 1 \\
K2  & --    &  4760  & 4710  &  4450$\pm$123  & 1 \\
K5  & --    &  4140  & 4210  &  4055$\pm$78   & 2 \\
K7  & --    &  3970  & 4020  &  3905$\pm$63   & 2 \\
M0  & --    &  3770  & 3900  &  3769$\pm$111  & 5 \\
M1  & 3705  &  3630  & 3720  &  3696$\pm$56   & 9 \\
M2  & 3560  &  3490  & 3560  &  3535$\pm$79   & 9 \\
M3  & 3415  &  3360  & 3410  &  3496$\pm$26   & 5 \\
M4  & 3270  &  3160  & 3190  &  3319$\pm$58  & 3 \\
M5  & 3125  &  2880  & 2980  &  3269$\pm$48  & 2
\enddata
\end{deluxetable}

\subsection{Surface gravity, stellar ages and masses}
The \logg--\teff\ plane, also known as the spectroscopic Hertzsprung-Russell diagram \citep[sHR;][]{langer14}, helps estimate stellar properties such as age and mass because \logg\ and \teff\ are proxies for those properties \citep[See Figure 9 of][]{sokal18}. {According to the evolutionary models of \cite{baraffe15}, over the mass range of 0.05--1.4~$M_{\odot}$, the \logg\ changes by about 0.5~dex between 1 to 10~Myr}. Therefore, two populations of different ages should occupy different parts of the sHR.

\begin{figure*}
    \centering
    \includegraphics[width=\textwidth]{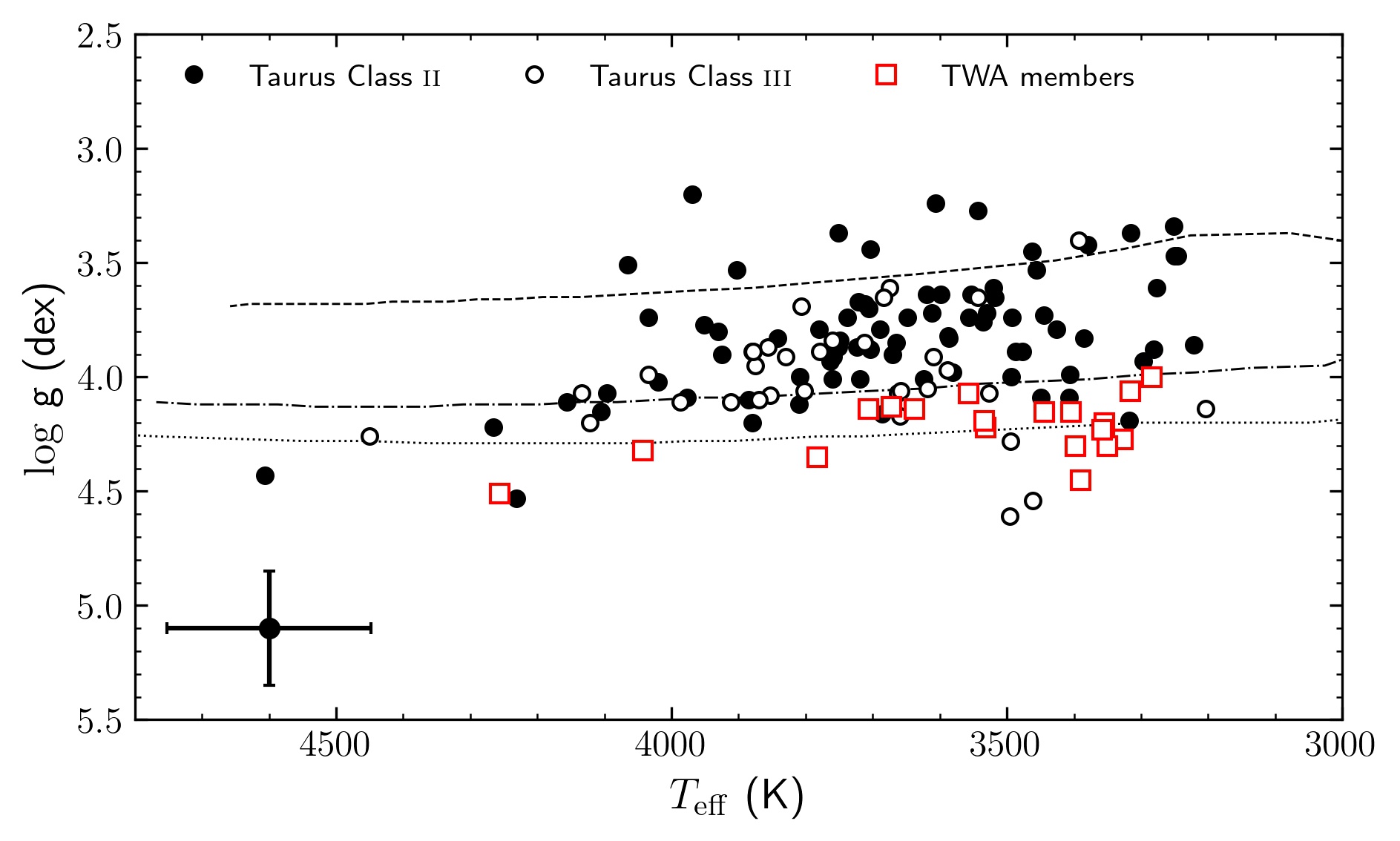}
    \caption{Spectroscopic Hertzsprung-Russell diagram. The solid and empty circles represent our determinations for Class~{\sc ii} and Class~{\sc iii} Taurus YSOs, while the empty squares are our determinations for the TWA members. These different YSO samples present an offset in \logg, which suggests an age ladder with the Taurus YSOs younger than their TWA counterparts. We also include the \citet{baraffe15} isochrones of 1 (dashed line), $\sim$5 (dash-dotted line) and 10~Myr (dotted line). The error bar represents our median fit uncertainties. \label{fig:temvsgra}}
    \end{figure*}

{With this in mind, we observed members of the TW Hydrae Association (TWA; \citealp{katsner97}). TWA is a nearby ($\sim$60 pc;  \citealp{zuckerman04}; \citealp{gaia18}) and young ($\sim$7--10 Myr; \citealp{ducourant14}; \citealp{herczeg15}; \citealp{sokal18}) group of stars, that serves as an evolved counterpart to our Taurus sample. 
We analyzed and determined the stellar parameters (Table~\ref{tab:twa}) for all 19 TWA objects in the same fashion as the Taurus sample.
In Figure~\ref{fig:temvsgra} we assemble the sHR of the Taurus and TWA YSOs, and compare them to the 1, $\sim$5, and 10~Myr isochrones of \cite{baraffe15}. }

\begin{deluxetable*}{rllccclllll}
\tablewidth{0pt}
 \tablecaption{Stellar parameters and basic information for 19 TWA members. The quality flag is the same as in Table~\ref{tab:res}. K magnitudes come from 2MASS \citep{cutri03} while the SpT references are indicated. The errors on the stellar parameters are the total (fit + systematic) uncertainties.\label{tab:twa}}
  
\tablehead{\colhead{2MASS ID} & \colhead{Name} & \colhead{SpT} & \colhead{Ref.} & \colhead{K} & \colhead{flag} & \colhead{\teff} & \colhead{\logg} & \colhead{B} & \colhead{\vsini} & \colhead{$r_{\rm K}$} \\
\colhead{} & \colhead{} &\colhead{} &\colhead{} & \colhead{(mag)} &\colhead{} & \colhead{(K)} & \colhead{(dex)} & \colhead{(kG)} & \colhead{(km s$^{-1}$)} & \colhead{} }
\startdata
J11015191-3442170 &  TWA 1 &  M0.5 & 1 & 7.3 &  0 & 3783$\pm$108 & 4.35$\pm$0.18 & 2.75$\pm$0.30 &  8.4$\pm$2.0 & 0.53$\pm$0.05\\
J11091380-3001398 &  TWA 2 &  M2.2 & 1 & 6.7 &  0 & 3558$\pm$ 92 & 4.07$\pm$0.16 &       $<$1.38 & 15.9$\pm$1.9 & 0.06$\pm$0.02\\
J11102788-3731520 & TWA 3A &  M4.1 & 1 & 6.8 &  1 & 3285$\pm$ 75 & 4.00$\pm$0.13 &       $<$0.97 & 11.9$\pm$1.7 & 0.29$\pm$0.01\\
              --  & TWA 3B &  M4.0 & 1 &  -- &  1 & 3355$\pm$ 75 & 4.20$\pm$0.13 &       $<$1.78 & 15.8$\pm$1.7 & 0.16$\pm$0.01\\
J11220530-2446393 &  TWA 4 &  K6.0 & 1 & 5.6 &  0 & 4257$\pm$103 & 4.51$\pm$0.21 &       $<$1.10 & 10.2$\pm$2.1 & 0.31$\pm$0.04\\
J10423011-3340162 &  TWA 7 &  M3.2 & 1 & 6.9 &  0 & 3328$\pm$ 80 & 4.27$\pm$0.13 & 2.19$\pm$0.26 &  7.4$\pm$1.7 & 0.10$\pm$0.01\\
J11324124-2651559 & TWA 8A &  M2.9 & 1 & 7.4 &  0 & 3398$\pm$ 75 & 4.30$\pm$0.13 & 2.86$\pm$0.26 &  7.5$\pm$1.7 & 0.13$\pm$0.01\\
J11482422-3728491 & TWA 9A &  K6.0 & 1 & 7.8 &  0 & 4043$\pm$ 94 & 4.32$\pm$0.17 & 2.29$\pm$0.30 & 12.0$\pm$2.0 & 0.17$\pm$0.03\\
J11482373-3728485 & TWA 9B &  M3.4 & 1 & 9.2 &  0 & 3351$\pm$ 75 & 4.30$\pm$0.14 &       $<$1.30 & 10.9$\pm$1.7 & 0.09$\pm$0.01\\
J12350424-4136385 & TWA 10 &  M3.0 & 2 & 8.2 &  1 & 3358$\pm$ 75 & 4.23$\pm$0.13 & 2.54$\pm$0.26 &  9.5$\pm$1.7 & 0.10$\pm$0.01\\
J12360055-3952156 &TWA 11B &  M2.5 & 3 & 8.3 &  0 & 3532$\pm$115 & 4.22$\pm$0.18 & 2.23$\pm$0.34 & 14.8$\pm$2.1 & 0.02$\pm$0.02\\
12354893-3950245, &TWA 11C &  M4.5 & 2 & 8.9 &  1 & 3391$\pm$122 & 4.45$\pm$0.22 &       $<$1.98 & 22.8$\pm$2.3 & 0.36$\pm$0.04\\
J11210549-3845163 & TWA 12 & M2.75 & 2 & 8.1 &  0 & 3534$\pm$104 & 4.19$\pm$0.17 & 2.67$\pm$0.32 & 19.7$\pm$2.0 & 0.02$\pm$0.02\\
J11211723-3446454 &TWA 13A &  M1.1 & 1 & 7.5 &  0 & 3638$\pm$ 96 & 4.14$\pm$0.16 & 2.12$\pm$0.29 & 14.2$\pm$1.9 & 0.04$\pm$0.02\\
J11211745-3446497 &TWA 13B &  M1.0 & 1 & 7.5 &  0 & 3673$\pm$ 93 & 4.13$\pm$0.16 & 1.71$\pm$0.30 & 13.7$\pm$1.9 & 0.04$\pm$0.02\\
J12345629-4538075 & TWA 16 &  M3.0 & 2 & 8.1 &  0 & 3445$\pm$ 91 & 4.15$\pm$0.16 & 1.85$\pm$0.28 & 12.4$\pm$1.9 & 0.07$\pm$0.02\\
J12072738-3247002 & TWA 23 &  M3.5 & 1 & 7.8 &  1 & 3405$\pm$101 & 4.15$\pm$0.18 &       $<$1.42 & 19.1$\pm$1.9 & 0.16$\pm$0.03\\
J12153072-3948426 & TWA 25 &  M0.5 & 1 & 7.3 &  0 & 3707$\pm$ 99 & 4.14$\pm$0.17 & 2.21$\pm$0.29 & 15.0$\pm$1.9 & 0.06$\pm$0.03\\
J10120908-3124451 & TWA 39 &  M4.0 & 4 & 8.0 &  1 & 3316$\pm$ 88 & 4.06$\pm$0.18 &       $<$1.34 & 18.0$\pm$2.0 & 0.28$\pm$0.03\\
\enddata
\tablereferences{ (1) \citet{herczeg14}; 
(2) \citet{luhman17};
(3) \citet{webb99}; 
(4)  \citet{riedel14}}
\end{deluxetable*} 

As expected, the Taurus and TWA YSOs populate different parts of the sHR, exhibiting an offset in \logg. Most of the Taurus sample is located between the 1 and 5~Myr isochrones while the TWA members are located below the 5~Myr isochrone. We found a mean \logg\ (and error on the mean) of 3.87$\pm$0.03~dex and 4.22$\pm$0.03~dex, for Taurus and TWA, respectively.

We applied to the \logg\ distributions of Taurus and TWA a Kolmogorov-Smirnov (KS) test\footnote{We have used the KS\_2samp function, which is part of the python package Scipy.stats}. The KS test explores the null hypothesis that two samples are drawn from the same distribution. If the probability value ($\mathcal{P(\rm KS)}$) obtained from this test is higher than a certain threshold, often set at 1\%, 5\%, or 10\%, we cannot reject the null hypothesis. {We found a $\mathcal{P(\rm KS)} \sim 10^{-8}$ that the \logg\ distributions of Taurus and TWA come from the same parent distribution, validating that precise \logg\ values can provide meaningful age proxies for populations of stars.}

{For the mean \teff\ and \logg\ values, and the set of isochrones from \cite{baraffe15}, we can make a rough estimate of the age of Taurus and TWA. Our calculations lead to a Taurus mean age of $\sim$2.5~Myr and a TWA mean age of $\sim$10~Myr. 
However, the dispersion within these groups is larger than the differences between them.
The parameters of the least evolved TWA members are more similar to the more evolved members of Taurus than to the most evolved members of TWA.}

\subsection{Class {\sc ii} vs. Class~{\sc iii} YSOs}
Based on the shape of their spectral energy distributions, \cite*{lada87} divided YSOs into three different morphological classes ({\sc i, ii,} and {\sc  iii}) using the spectral index $n = d\log (\lambda F_\lambda)/d\log \lambda$, and suggested that such classes represent an evolutionary sequence in the formation of low-mass stars. Since then, the sequence has been extended to Class 0 YSOs embedded in an infalling cold dust envelope \citep{andre93}, flat spectrum YSOs \citep{greene94} between Class {\sc i} and {\sc ii}, and post-Class~{\sc iii} transition disk objects without the cold dust corresponding to flux from 5-20~$\mu$m \citep{lada06}. 
If the classification of YSOs by their SED slope, as suggested, represents an evolutionary scheme, the populations of YSOs would differ in terms of stellar properties, such as \logg, given that stars undergo gravitational contraction while they evolve towards the main-sequence \citep{herbig62}. 

To test this, we collected YSO classifications from the literature for 84 Class~{\sc ii} and 35 Class~{\sc iii}  (78 and 32 with good or acceptable parameters, respectively) objects in our sample \citep{kenyon95,luhman2010,rebull2010,esplin2014,kraus17}. We used a KS test to quantify if the stellar parameters determined for the Class~{\sc ii} and Class~{\sc iii} stars are statistically different.

{We found a $\mathcal{P}$(KS) of 32\%, 9\%, and 6\% that the Class~{\sc ii} and Class~{\sc iii} objects come from the same parent distributions of \teff, \vsini, and B respectively.\footnote{We considered just the 41 B determinations in this test, not the limits.} This probability dropped to values lower than 1\% for \logg\ and $r_{\rm K}$. These probability values indicate that the Class~{\sc ii} and Class~{\sc iii} YSOs of our sample are statistically indistinguishable in terms of \teff, marginally indistinguishable in terms of \vsini\ and B, but different in terms of $r_{\rm K}$ and \logg.}

The result of the KS tests for \teff\ and $r_{\rm K}$ are not surprising. The distribution of temperatures (SpT) reflects the initial mass function of Taurus.
Also, Class~{\sc ii} YSOs host disks and are actively accreting, which correlate with high levels of veiling, while Class~{\sc iii} objects have lost their disks and show little or no veiling. The veiling values we measure for Class~{\sc iii} objects is closer to zero than what we find for Class~{\sc ii} YSOs, consistent with the expected behavior. 
{For \vsini, the $\mathcal{P}$(KS) of 9\% that we found is in agreement with the findings of \cite{nguyen09}, who found that low-mass accretors and non-accretors in Taurus have a $\mathcal{P}$(KS)~=~10\% of coming from the same \vsini\ parent distribution. }

\begin{figure}
    \centering
    \includegraphics[width=\columnwidth]{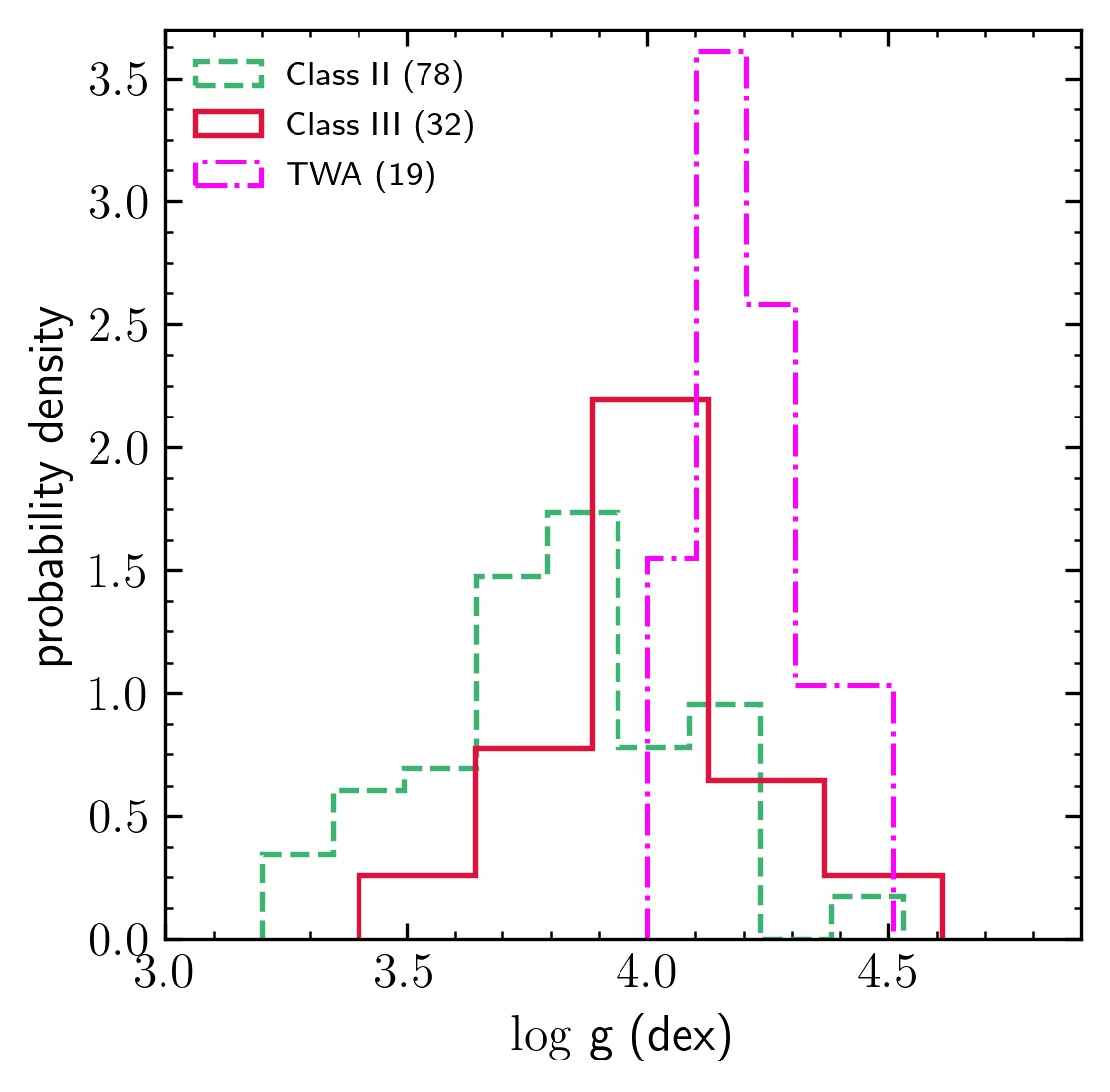}
    \caption{Probability densities of \logg\ for 78 Class~{\sc ii} (dashed) and 32 Class~{\sc iii} (solid) Taurus YSOs. We also included the \logg\ probability density of the TWA YSOs as a comparison. We computed a KS probability between the Class~{\sc ii} and Class~{\sc iii} of $\sim3\times10^{-3}$. If we compare the \logg\ distributions of the Taurus and TWA YSOs, we also found low KS probabilities of  $\sim1.5\times10^{-9}$ and $\sim7\times10^{-5}$, for Class~{\sc ii} and Class~{\sc iii}, respectively. These low probability values show that Taurus Class~{\sc ii} and Class~{\sc iii} YSOs, as well as the TWA objects, are different in terms of their \logg. \label{fig:proba}}
\end{figure}

Since \logg\ can be used as a proxy for age \citep{yao18}, the different distributions for Class~{\sc ii} and Class~{\sc iii} YSOs support the idea of an evolutionary sequence (see Figure~\ref{fig:proba}). We found a mean \logg\ of 3.83$\pm$0.03 and 4.06$\pm$0.04~dex for Class~{\sc ii} and Class~{\sc iii} Taurus objects, respectively. 
The most interesting sources in each \logg\ distribution are those in the wings, where evolutionary differences between the lowest \logg\ Class~{\sc iii} YSOs and the highest \logg\ Class~{\sc ii} YSOs are still not well understood.

\begin{figure}
    \centering
    \includegraphics[width=\columnwidth]{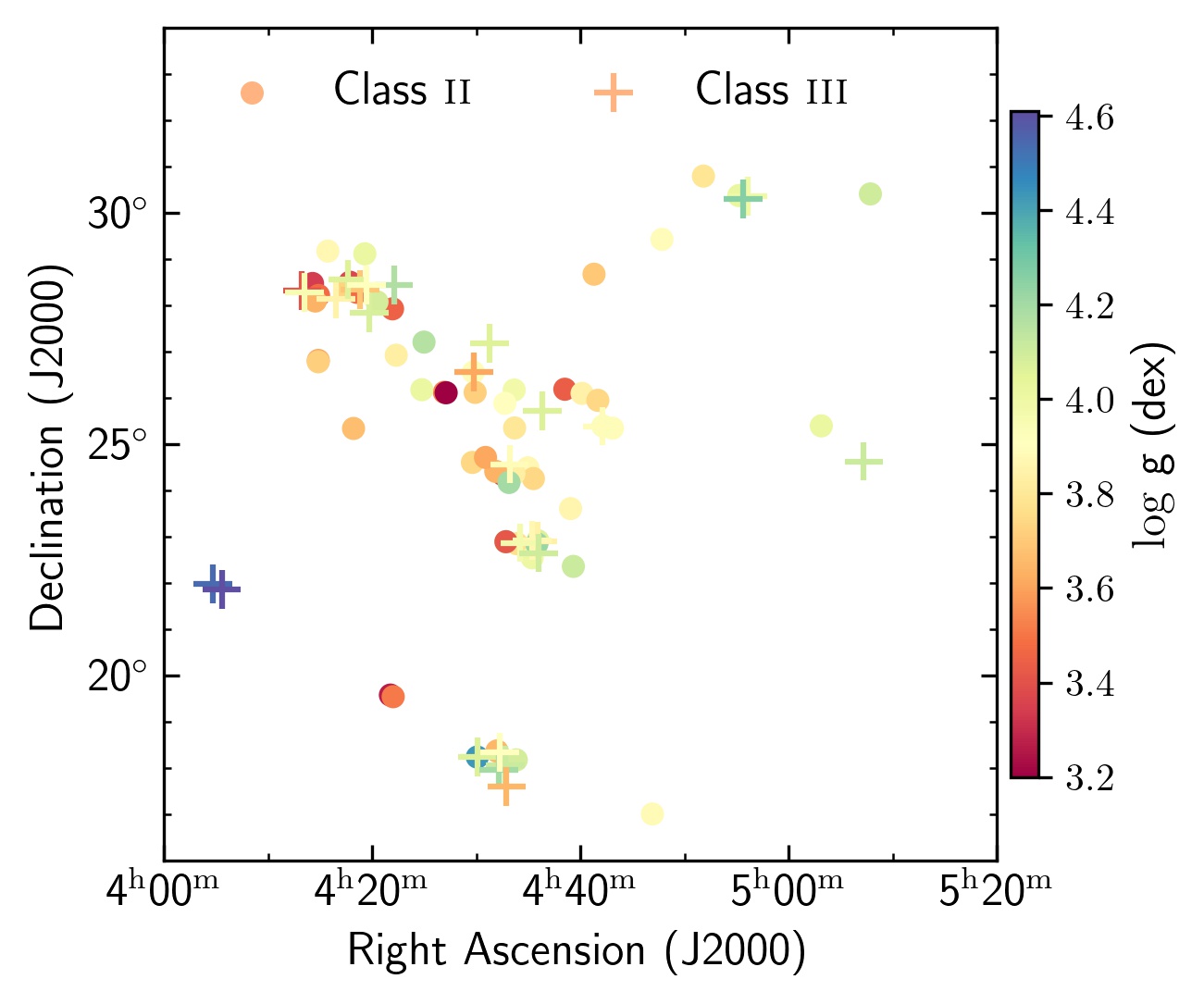}
    \caption{The spatial distribution of the Taurus sample color-coded by \logg. The 78 Class~{\sc ii} (circles) and 32 Class~{\sc iii} (crosses) YSOs are distributed similarly across the sky. There is no significant clustering of Class~{\sc ii} or Class~{\sc iii} objects in the sample we have studied.\label{fig:map}}
\end{figure}

Finally, the spatial distribution for the IGRINS Taurus sample (Figure~\ref{fig:map}) does not show significant clustering of Class~{\sc ii} or Class~{\sc iii} objects, nor a trend with \logg, suggesting that there is not a star formation gradient across the cloud. Yet, this is a 2-dimensional view of Taurus, and it will require the entire population of Taurus YSOs \citep[$>$400;][]{luhman2018} studied in the context of 3-dimensional substructures \citep{Krolikowski2021} before we can fully understand the spatial distribution in relation to the parameters we have determined.

\subsection{Important Considerations}
\label{sect:importance}
There are many ways to perform the parameter determination that we present here. 
As outlined above, there are a variety of models to choose from and possible approaches. 
We have shown that magnetic fields are a critical component of the parameter determination, and we allowed many of the YSO parameters to remain variables in the fit rather than fixing them to assumed values.
During our tests, we found that \logg\ was the most sensitive parameter to variations in our approach.
Additionally, we fit for all parameters simultaneously rather than taking an iterative approach to the fitting.
While deciding on the best method for parameter determination, by minimizing fit residuals and validating the results against the literature, we have identified some important considerations when performing YSO parameter determinations. \\

{\it Use of consistent \teff\ estimators}: IGRINS provides us with both the H- and K-band spectrum of our sources, but we have only employed the K-band in our analysis. 
There are temperature sensitive spectral features in the H-band (the OH region of \citealp{lopezval19}) that we initially included in our analysis.
The result of including this H-band region was an offset in \logg\ towards higher values, inconsistent with the evolutionary stage of the Taurus star-forming region. 
Using both the H- and K-band spectra meant that \logg\ came primarily from the CO region and \teff\ came from the OH region, which are about 0.7~$\mu$m apart.
Furthermore, \cite{gully17} found that \teff\ varies as a function of wavelength when a single temperature component was fit to the IGRINS spectrum of the spotted YSO LkCa~4. The temperature in the CO region of LkCa~4 is $\sim$700~K cooler than the value determined in the OH region (see Figure 3 of \citealp{gully17}). 
In this work, differences between the H- and K-band temperatures were compensated for by the remaining stellar parameters, which primarily impacted the \logg\ determination.
While the broad spectral grasp of IGRINS permits the analysis of additional features, it also probes different components of the YSO photosphere, which can produce conflicting measurements. 
We have mitigated these effects by fitting only the K-band regions of the Taurus YSOs.\\

{\it Degeneracy between $r_{\rm K}$ and \logg:} Another consideration was the degeneracy between \logg\ and $r_{\rm K}$. As seen in the diagonal distribution of $r_{\rm K}$ and \logg\ in the sub-panel of Figure~\ref{fig:corner}, higher values of $r_{\rm K}$ results in lower \logg\ values, and vice versa. This degeneracy results at a fixed \teff\ because $r_{\rm K}$ and \logg\ impact the CO lines in nearly in the same way, and the CO lines are the most \logg\ sensitive lines in our analysis. 
This degeneracy increases the uncertainties for each parameter and we concluded that \logg\ and $r_{\rm K}$ need to be determined simultaneously and from the same spectrum. 
The use of fixed $r_{\rm K}$ or \logg\ values from the literature will disproportionately impact the other parameter. \\ 

{\it The atomic data: } Our method relies on the comparison of observed and synthetic spectra, and the line list and the model atmosphere selected is a crucial component of the result. 
We initially determined the stellar parameters using the unmodified VALD line list expecting that the use of different spectral regions would average the inaccuracies of the atomic data. 
However, the resultant stellar parameters were offset from previous determinations in the literature.  
We then modified the oscillator strengths and the van der Wals constants of our spectral regions to the values of \cite{flores19}, which improved the stellar parameter determinations and the consistency with the literature. 
We used the \cite{flores19} modifications because they were obtained from spectra with a spectral resolution close to IGRINS and for the same spectral regions we have employed. \\

{\it Comparison with the literature: } 
The bright YSOs in this work are canonical members of the Taurus-Auriga star forming region and have numerous published parameters.
Many of the prior works in the literature \citep[e.g.,][]{johns-krull07,pecaut13,herczeg14} fixed parameters that we have left as variables in the fits.
Prior studies also analyzed shorter wavelength regions of the YSO spectra, which are dominated by the hotter photosphere.
Additionally, the variable nature of YSOs means that observations made at different epochs likely have different physical properties.
In comparing our results to the literature, we find that the observed spectra and methods of \cite{flores19} provide the most direct comparisons and consistency.

\section{Summary and conclusions}
{ We employed four different spectral regions of the infrared K-band and an iterative MCMC approach to determine \teff, \logg, B, and \vsini\ for a sample of canonical YSOs in the Taurus-Auriga star-forming region.}
Our analyses used a solar metallicity synthetic spectral grid of MARCS atmosphere models and the spectral synthesis code {\sc moogstokes}. Our observational data were obtained with the Immersion GRating INfrared Spectrometer (IGRINS). 

We were able to reliably determine \teff, \logg, and \vsini\ for 110 YSOs. Our ability to determine B is dependent on the \vsini\ and limited by the spectral resolution of IGRINS. We established that B values $>$1~kG or (\vsini/8)~kG are required to provide meaningful B-field detections, and that SNR$>$150 is desirable for overall parameter fitting. 
Measurable B-fields were found for { 41} of the YSOs, and upper limits were identified for the remaining objects.

{The \teff\ we derived agrees with previous temperature scales for spectral type M YSOs, but it is systematically cooler for K stars. 
This could be a result of the models used in this analysis compared to others or the wavelength regions studied, which are more sensitive to the cool spotted regions of hotter YSOs. 
Our results show that the Class~{\sc ii} and Class~{\sc iii} objects are statistically similar in terms of \teff, {marginally similar in \vsini\ and B, and different in \logg\ and r$_{\rm K}$}. The differences found in the distributions of \logg\ and r$_{\rm K}$ correspond to the expected behavior if Class~{\sc iii} are more evolved than Class~{\sc ii} objects.
We found mean \logg\ values of 3.83$\pm$0.03, 4.06$\pm$0.04, and 4.22$\pm$0.03 dex for the Taurus Class~{\sc ii} and Class~{\sc iii} objects and 19 TWA members, respectively. These differences are statistically significant and affirms the age gradient between these young star populations.}


\acknowledgements
{We thank the anonymous referees for suggestions that have improved our methodology and scientific results, as well as for inspiring the summary of important considerations in Section~\ref{sect:importance}.}  
This work used the Immersion Grating Infrared Spectrograph (IGRINS) that was developed under a 
collaboration between the University of Texas at Austin and the Korea Astronomy and Space Science 
Institute (KASI) with the financial support of the US National Science Foundation under grant 
AST-1229522 and AST-1702267, of the University of Texas at Austin, and of the Korean GMT Project of 
KASI. These results made use of the Lowell Discovery Telescope. Lowell is a 
private, non-profit institution dedicated to astrophysical research and public appreciation of 
astronomy and operates the LDT in partnership with Boston University, the University of Maryland, the 
University of Toledo, Northern Arizona University and Yale University. This paper includes data taken 
at The McDonald Observatory of The University of Texas at Austin. Based on observations obtained at the Gemini Observatory, which is operated by the Association of Universities for Research in Astronomy, Inc., under a cooperative agreement with the NSF on behalf of the Gemini partnership: the National Science Foundation (United States), the National Research Council (Canada), CONICYT (Chile), Ministerio de Ciencia, Tecnolog\'{i}a e Innovaci\'{o}n Productiva (Argentina), and Minist\'{e}rio da Ci\^{e}ncia, Tecnologia e Inova\c{c}\~{a}o (Brazil). 
This material is based upon work supported by the National Science Foundation under Grant No.\ AST-1908892 to G.\ Mace.

\software{IGRINS pipeline package \citep{lee17}, {\sc moogstokes} \citep{deen13}, MARCS models \citep{marcs}, {\sc zbarycorr} \citep{wright2014}, astroquery \citep{astroquery}, emcee \citep{emcee}, matplotlib \citep{matplotlib}, numpy \citep{numpy}, pandas \citep{pandas}, astropy \citep{astropy, astropy2}, scipy \citep{scipy}}

\bibliography{taurus_R4}

\begin{thebibliography}{}
\expandafter\ifx\csname natexlab\endcsname\relax\def\natexlab#1{#1}\fi
\providecommand{\url}[1]{\href{#1}{#1}}

\bibitem[{{Adams} {et~al.}(1987){Adams}, {Lada}, \& {Shu}}]{lada87}
{Adams}, F.~C., {Lada}, C.~J., \& {Shu}, F.~H. 1987, \apj, 312, 788

\bibitem[{{ALMA Partnership} {et~al.}(2015){ALMA Partnership}, {Brogan},
  {P{\'e}rez}, {Hunter}, {Dent}, {Hales}, {Hills}, {Corder}, {Fomalont},
  {Vlahakis}, {Asaki}, {Barkats}, {Hirota}, {Hodge}, {Impellizzeri}, {Kneissl},
  {Liuzzo}, {Lucas}, {Marcelino}, {Matsushita}, {Nakanishi}, {Phillips},
  {Richards}, {Toledo}, {Aladro}, {Broguiere}, {Cortes}, {Cortes}, {Espada},
  {Galarza}, {Garcia-Appadoo}, {Guzman-Ramirez}, {Humphreys}, {Jung}, {Kameno},
  {Laing}, {Leon}, {Marconi}, {Mignano}, {Nikolic}, {Nyman}, {Radiszcz},
  {Remijan}, {Rod{\'o}n}, {Sawada}, {Takahashi}, {Tilanus}, {Vila Vilaro},
  {Watson}, {Wiklind}, {Akiyama}, {Chapillon}, {de Gregorio-Monsalvo}, {Di
  Francesco}, {Gueth}, {Kawamura}, {Lee}, {Nguyen Luong}, {Mangum}, {Pietu},
  {Sanhueza}, {Saigo}, {Takakuwa}, {Ubach}, {van Kempen}, {Wootten},
  {Castro-Carrizo}, {Francke}, {Gallardo}, {Garcia}, {Gonzalez}, {Hill},
  {Kaminski}, {Kurono}, {Liu}, {Lopez}, {Morales}, {Plarre}, {Schieven},
  {Testi}, {Videla}, {Villard}, {Andreani}, {Hibbard}, \& {Tatematsu}}]{alma15}
{ALMA Partnership}, {Brogan}, C.~L., {P{\'e}rez}, L.~M., {et~al.} 2015, \apjl,
  808, L3

\bibitem[{{Andre} {et~al.}(1993){Andre}, {Ward-Thompson}, \&
  {Barsony}}]{andre93}
{Andre}, P., {Ward-Thompson}, D., \& {Barsony}, M. 1993, \apj, 406, 122

\bibitem[{{Astropy Collaboration} {et~al.}(2013){Astropy Collaboration},
  {Robitaille}, {Tollerud}, {Greenfield}, {Droettboom}, {Bray}, {Aldcroft},
  {Davis}, {Ginsburg}, {Price-Whelan}, {Kerzendorf}, {Conley}, {Crighton},
  {Barbary}, {Muna}, {Ferguson}, {Grollier}, {Parikh}, {Nair}, {Unther},
  {Deil}, {Woillez}, {Conseil}, {Kramer}, {Turner}, {Singer}, {Fox}, {Weaver},
  {Zabalza}, {Edwards}, {Azalee Bostroem}, {Burke}, {Casey}, {Crawford},
  {Dencheva}, {Ely}, {Jenness}, {Labrie}, {Lim}, {Pierfederici}, {Pontzen},
  {Ptak}, {Refsdal}, {Servillat}, \& {Streicher}}]{astropy}
{Astropy Collaboration}, {Robitaille}, T.~P., {Tollerud}, E.~J., {et~al.} 2013,
  \aap, 558, A33

\bibitem[{{Astropy Collaboration} {et~al.}(2018){Astropy Collaboration},
  {Price-Whelan}, {Sip{\H{o}}cz}, {G{\"u}nther}, {Lim}, {Crawford}, {Conseil},
  {Shupe}, {Craig}, {Dencheva}, {Ginsburg}, {VanderPlas}, {Bradley},
  {P{\'e}rez-Su{\'a}rez}, {de Val-Borro}, {Aldcroft}, {Cruz}, {Robitaille},
  {Tollerud}, {Ardelean}, {Babej}, {Bach}, {Bachetti}, {Bakanov}, {Bamford},
  {Barentsen}, {Barmby}, {Baumbach}, {Berry}, {Biscani}, {Boquien}, {Bostroem},
  {Bouma}, {Brammer}, {Bray}, {Breytenbach}, {Buddelmeijer}, {Burke},
  {Calderone}, {Cano Rodr{\'\i}guez}, {Cara}, {Cardoso}, {Cheedella}, {Copin},
  {Corrales}, {Crichton}, {D'Avella}, {Deil}, {Depagne}, {Dietrich}, {Donath},
  {Droettboom}, {Earl}, {Erben}, {Fabbro}, {Ferreira}, {Finethy}, {Fox},
  {Garrison}, {Gibbons}, {Goldstein}, {Gommers}, {Greco}, {Greenfield},
  {Groener}, {Grollier}, {Hagen}, {Hirst}, {Homeier}, {Horton}, {Hosseinzadeh},
  {Hu}, {Hunkeler}, {Ivezi{\'c}}, {Jain}, {Jenness}, {Kanarek}, {Kendrew},
  {Kern}, {Kerzendorf}, {Khvalko}, {King}, {Kirkby}, {Kulkarni}, {Kumar},
  {Lee}, {Lenz}, {Littlefair}, {Ma}, {Macleod}, {Mastropietro}, {McCully},
  {Montagnac}, {Morris}, {Mueller}, {Mumford}, {Muna}, {Murphy}, {Nelson},
  {Nguyen}, {Ninan}, {N{\"o}the}, {Ogaz}, {Oh}, {Parejko}, {Parley}, {Pascual},
  {Patil}, {Patil}, {Plunkett}, {Prochaska}, {Rastogi}, {Reddy Janga},
  {Sabater}, {Sakurikar}, {Seifert}, {Sherbert}, {Sherwood-Taylor}, {Shih},
  {Sick}, {Silbiger}, {Singanamalla}, {Singer}, {Sladen}, {Sooley},
  {Sornarajah}, {Streicher}, {Teuben}, {Thomas}, {Tremblay}, {Turner},
  {Terr{\'o}n}, {van Kerkwijk}, {de la Vega}, {Watkins}, {Weaver}, {Whitmore},
  {Woillez}, {Zabalza}, \& {Astropy Contributors}}]{astropy2}
{Astropy Collaboration}, {Price-Whelan}, A.~M., {Sip{\H{o}}cz}, B.~M., {et~al.}
  2018, \aj, 156, 123

\bibitem[{{Attridge} \& {Herbst}(1992)}]{attridge92}
{Attridge}, J.~M., \& {Herbst}, W. 1992, \apjl, 398, L61

\bibitem[{{Baraffe} {et~al.}(2015){Baraffe}, {Homeier}, {Allard}, \&
  {Chabrier}}]{baraffe15}
{Baraffe}, I., {Homeier}, D., {Allard}, F., \& {Chabrier}, G. 2015, \aap, 577,
  A42

\bibitem[{{Barenfeld} {et~al.}(2017){Barenfeld}, {Carpenter}, {Sargent},
  {Isella}, \& {Ricci}}]{barenfeld17}
{Barenfeld}, S.~A., {Carpenter}, J.~M., {Sargent}, A.~I., {Isella}, A., \&
  {Ricci}, L. 2017, \apj, 851, 85

\bibitem[{{Baruteau} {et~al.}(2014){Baruteau}, {Crida}, {Paardekooper},
  {Masset}, {Guilet}, {Bitsch}, {Nelson}, {Kley}, \& {Papaloizou}}]{baruteau14}
{Baruteau}, C., {Crida}, A., {Paardekooper}, S.~J., {et~al.} 2014, in
  Protostars and Planets VI, ed. H.~{Beuther}, R.~S. {Klessen}, C.~P.
  {Dullemond}, \& T.~{Henning}, 667

\bibitem[{{Basri} \& {Batalha}(1990)}]{basri90}
{Basri}, G., \& {Batalha}, C. 1990, \apj, 363, 654

\bibitem[{{Basri} \& {Bertout}(1989)}]{basri89}
{Basri}, G., \& {Bertout}, C. 1989, \apj, 341, 340

\bibitem[{{Bean} {et~al.}(2006){Bean}, {Sneden}, {Hauschildt}, {Johns-Krull},
  \& {Benedict}}]{bean06}
{Bean}, J.~L., {Sneden}, C., {Hauschildt}, P.~H., {Johns-Krull}, C.~M., \&
  {Benedict}, G.~F. 2006, \apj, 652, 1604

\bibitem[{{Boccaletti} {et~al.}(2020){Boccaletti}, {Di Folco}, {Pantin},
  {Dutrey}, {Guilloteau}, {Tang}, {Pi{\'e}tu}, {Habart}, {Milli}, {Beck}, \&
  {Maire}}]{boccaletti2020}
{Boccaletti}, A., {Di Folco}, E., {Pantin}, E., {et~al.} 2020, \aap, 637, L5

\bibitem[{{Bouvier} {et~al.}(2014){Bouvier}, {Matt}, {Mohanty}, {Scholz},
  {Stassun}, \& {Zanni}}]{bouvier14}
{Bouvier}, J., {Matt}, S.~P., {Mohanty}, S., {et~al.} 2014, in Protostars and
  Planets VI, ed. H.~{Beuther}, R.~S. {Klessen}, C.~P. {Dullemond}, \&
  T.~{Henning}, 433

\bibitem[{{Bouvier} {et~al.}(1992){Bouvier}, {Tessier}, \&
  {Cabrit}}]{bouvier92}
{Bouvier}, J., {Tessier}, E., \& {Cabrit}, S. 1992, \aap, 261, 451

\bibitem[{{Calvet} {et~al.}(1984){Calvet}, {Basri}, \& {Kuhi}}]{calvet87}
{Calvet}, N., {Basri}, G., \& {Kuhi}, L.~V. 1984, \apj, 277, 725

\bibitem[{{Camenzind}(1990)}]{camenzind90}
{Camenzind}, M. 1990, Reviews in Modern Astronomy, 3, 234

\bibitem[{{Casagrande} {et~al.}(2008){Casagrande}, {Flynn}, \&
  {Bessell}}]{casagrande08}
{Casagrande}, L., {Flynn}, C., \& {Bessell}, M. 2008, \mnras, 389, 585

\bibitem[{{Cieza} {et~al.}(2005){Cieza}, {Kessler-Silacci}, {Jaffe}, {Harvey},
  \& {Evans}}]{cieza05}
{Cieza}, L.~A., {Kessler-Silacci}, J.~E., {Jaffe}, D.~T., {Harvey}, P.~M., \&
  {Evans}, Neal~J., I. 2005, \apj, 635, 422

\bibitem[{{Correia} {et~al.}(2006){Correia}, {Zinnecker}, {Ratzka}, \&
  {Sterzik}}]{correia06}
{Correia}, S., {Zinnecker}, H., {Ratzka}, T., \& {Sterzik}, M.~F. 2006, \aap,
  459, 909

\bibitem[{{Cottaar} {et~al.}(2014){Cottaar}, {Covey}, {Meyer}, {Nidever},
  {Stassun}, {Foster}, {Tan}, {Chojnowski}, {da Rio}, {Flaherty}, {Frinchaboy},
  {Skrutskie}, {Majewski}, {Wilson}, \& {Zasowski}}]{cottar14}
{Cottaar}, M., {Covey}, K.~R., {Meyer}, M.~R., {et~al.} 2014, The Astrophysical
  Journal, 794, 125

\bibitem[{{Crockett} {et~al.}(2012){Crockett}, {Mahmud}, {Prato},
  {Johns-Krull}, {Jaffe}, {Hartigan}, \& {Beichman}}]{crockett12}
{Crockett}, C.~J., {Mahmud}, N.~I., {Prato}, L., {et~al.} 2012, \apj, 761, 164

\bibitem[{{Cutri} {et~al.}(2003){Cutri}, {Skrutskie}, {van Dyk}, {Beichman},
  {Carpenter}, {Chester}, {Cambresy}, {Evans}, {Fowler}, {Gizis}, {Howard},
  {Huchra}, {Jarrett}, {Kopan}, {Kirkpatrick}, {Light}, {Marsh}, {McCallon},
  {Schneider}, {Stiening}, {Sykes}, {Weinberg}, {Wheaton}, {Wheelock}, \&
  {Zacarias}}]{cutri03}
{Cutri}, R.~M., {Skrutskie}, M.~F., {van Dyk}, S., {et~al.} 2003, VizieR Online
  Data Catalog, II/246

\bibitem[{{Deen}(2013)}]{deen13}
{Deen}, C.~P. 2013, \aj, 146, 51

\bibitem[{{Doppmann} {et~al.}(2003){Doppmann}, {Jaffe}, \&
  {White}}]{doppmann03-2}
{Doppmann}, G.~W., {Jaffe}, D.~T., \& {White}, R.~J. 2003, \aj, 126, 3043

\bibitem[{{D'Orazi} {et~al.}(2011){D'Orazi}, {Biazzo}, \& {Randich}}]{dorazi11}
{D'Orazi}, V., {Biazzo}, K., \& {Randich}, S. 2011, \aap, 526, A103

\bibitem[{{Duch{\^e}ne} {et~al.}(1999){Duch{\^e}ne}, {Monin}, {Bouvier}, \&
  {M{\'e}nard}}]{duchene99}
{Duch{\^e}ne}, G., {Monin}, J.~L., {Bouvier}, J., \& {M{\'e}nard}, F. 1999,
  \aap, 351, 954

\bibitem[{{Ducourant} {et~al.}(2014){Ducourant}, {Teixeira}, {Galli}, {Le
  Campion}, {Krone-Martins}, {Zuckerman}, {Chauvin}, \& {Song}}]{ducourant14}
{Ducourant}, C., {Teixeira}, R., {Galli}, P.~A.~B., {et~al.} 2014, \aap, 563,
  A121

\bibitem[{{Dyck} {et~al.}(1982){Dyck}, {Simon}, \& {Zuckerman}}]{dyck82}
{Dyck}, H.~M., {Simon}, T., \& {Zuckerman}, B. 1982, \apjl, 255, L103

\bibitem[{{Esplin} {et~al.}(2014){Esplin}, {Luhman}, \& {Mamajek}}]{esplin2014}
{Esplin}, T.~L., {Luhman}, K.~L., \& {Mamajek}, E.~E. 2014, \apj, 784, 126

\bibitem[{{Fischer} {et~al.}(2011){Fischer}, {Edwards}, {Hillenbrand}, \&
  {Kwan}}]{fischer11}
{Fischer}, W., {Edwards}, S., {Hillenbrand}, L., \& {Kwan}, J. 2011, \apj, 730,
  73

\bibitem[{{Flores} {et~al.}(2019){Flores}, {Connelley}, {Reipurth}, \&
  {Boogert}}]{flores19}
{Flores}, C., {Connelley}, M.~S., {Reipurth}, B., \& {Boogert}, A. 2019, \apj,
  882, 75

\bibitem[{{Flores} {et~al.}(2020){Flores}, {Reipurth}, \&
  {Connelley}}]{flores20}
{Flores}, C., {Reipurth}, B., \& {Connelley}, M.~S. 2020, arXiv e-prints,
  arXiv:2006.10139

\bibitem[{{Foreman-Mackey} {et~al.}(2013){Foreman-Mackey}, {Hogg}, {Lang}, \&
  {Goodman}}]{emcee}
{Foreman-Mackey}, D., {Hogg}, D.~W., {Lang}, D., \& {Goodman}, J. 2013, \pasp,
  125, 306

\bibitem[{{Gahm} {et~al.}(2008){Gahm}, {Walter}, {Stempels}, {Petrov}, \&
  {Herczeg}}]{gahm08}
{Gahm}, G.~F., {Walter}, F.~M., {Stempels}, H.~C., {Petrov}, P.~P., \&
  {Herczeg}, G.~J. 2008, \aap, 482, L35

\bibitem[{{Gaia Collaboration} {et~al.}(2018){Gaia Collaboration}, {Brown},
  {Vallenari}, {Prusti}, {de Bruijne}, {Babusiaux}, {Bailer-Jones}, {Biermann},
  {Evans}, {Eyer}, \& et~al.}]{gaia18}
{Gaia Collaboration}, {Brown}, A.~G.~A., {Vallenari}, A., {et~al.} 2018, \aap,
  616, A1

\bibitem[{{Galli} {et~al.}(2015){Galli}, {Bertout}, {Teixeira}, \&
  {Ducourant}}]{galli15}
{Galli}, P.~A.~B., {Bertout}, C., {Teixeira}, R., \& {Ducourant}, C. 2015,
  \aap, 580, A26

\bibitem[{{Galli} {et~al.}(2018){Galli}, {Loinard}, {Ortiz-L{\'e}on},
  {Kounkel}, {Dzib}, {Mioduszewski}, {Rodr{\'\i}guez}, {Hartmann}, {Teixeira},
  {Torres}, {Rivera}, {Boden}, {Evans}, {Brice{\~n}o}, {Tobin}, \&
  {Heyer}}]{galli18}
{Galli}, P. A.~B., {Loinard}, L., {Ortiz-L{\'e}on}, G.~N., {et~al.} 2018, \apj,
  859, 33

\bibitem[{{Gennaro} {et~al.}(2012){Gennaro}, {Prada Moroni}, \&
  {Tognelli}}]{gennaro12}
{Gennaro}, M., {Prada Moroni}, P.~G., \& {Tognelli}, E. 2012, \mnras, 420, 986

\bibitem[{{Ghez} {et~al.}(1993){Ghez}, {Neugebauer}, \& {Matthews}}]{ghez93}
{Ghez}, A.~M., {Neugebauer}, G., \& {Matthews}, K. 1993, \aj, 106, 2005

\bibitem[{{Ginsburg} {et~al.}(2019){Ginsburg}, {Sip{\H o}cz}, {Brasseur},
  {Cowperthwaite}, {Craig}, {Deil}, {Guillochon}, {Guzman}, {Liedtke}, {Lian
  Lim}, {Lockhart}, {Mommert}, {Morris}, {Norman}, {Parikh}, {Persson},
  {Robitaille}, {Segovia}, {Singer}, {Tollerud}, {de Val-Borro}, {Valtchanov},
  {Woillez}, {The Astroquery collaboration}, \& {a subset of the astropy
  collaboration}}]{astroquery}
{Ginsburg}, A., {Sip{\H o}cz}, B.~M., {Brasseur}, C.~E., {et~al.} 2019, \aj,
  157, 98

\bibitem[{{Gray}(2005)}]{gray05}
{Gray}, D.~F. 2005, {The Observation and Analysis of Stellar Photospheres}
  (Cambridge University Press), doi:10.1017/CBO9781316036570

\bibitem[{{Greene} {et~al.}(1994){Greene}, {Wilking}, {Andre}, {Young}, \&
  {Lada}}]{greene94}
{Greene}, T.~P., {Wilking}, B.~A., {Andre}, P., {Young}, E.~T., \& {Lada},
  C.~J. 1994, \apj, 434, 614

\bibitem[{Gully-Santiago {et~al.}(2010)Gully-Santiago, Wang, Deen, Kelly,
  Greene, Bacon, \& Jaffe}]{gully10}
Gully-Santiago, M., Wang, W., Deen, C., {et~al.} 2010, in Modern Technologies
  in Space- and Ground-based Telescopes and Instrumentation, ed.
  E.~Atad-Ettedgui \& D.~Lemke, Vol. 7739, International Society for Optics and
  Photonics (SPIE), 1353 -- 1359.
\newblock \url{https://doi.org/10.1117/12.857568}

\bibitem[{{Gully-Santiago} {et~al.}(2017){Gully-Santiago}, {Herczeg},
  {Czekala}, {Somers}, {Grankin}, {Covey}, {Donati}, {Alencar}, {Hussain},
  {Shappee}, {Mace}, {Lee}, {Holoien}, {Jose}, \& {Liu}}]{gully17}
{Gully-Santiago}, M.~A., {Herczeg}, G.~J., {Czekala}, I., {et~al.} 2017, \apj,
  836, 200

\bibitem[{{Gustafsson} {et~al.}(2008){Gustafsson}, {Edvardsson}, {Eriksson},
  {J{\o}rgensen}, {Nordlund}, \& {Plez}}]{marcs}
{Gustafsson}, B., {Edvardsson}, B., {Eriksson}, K., {et~al.} 2008, \aap, 486,
  951

\bibitem[{{Haas} {et~al.}(1990){Haas}, {Leinert}, \& {Zinnecker}}]{haas90}
{Haas}, M., {Leinert}, C., \& {Zinnecker}, H. 1990, \aap, 230, L1

\bibitem[{{Haisch} {et~al.}(2001){Haisch}, {Lada}, \& {Lada}}]{haisch01}
{Haisch}, Karl~E., J., {Lada}, E.~A., \& {Lada}, C.~J. 2001, \apjl, 553, L153

\bibitem[{Harris {et~al.}(2020)Harris, Millman, van~der Walt, Gommers,
  Virtanen, Cournapeau, Wieser, Taylor, Berg, Smith, Kern, Picus, Hoyer, van
  Kerkwijk, Brett, Haldane, Fernández~del Río, Wiebe, Peterson,
  Gérard-Marchant, Sheppard, Reddy, Weckesser, Abbasi, Gohlke, \&
  Oliphant}]{numpy}
Harris, C.~R., Millman, K.~J., van~der Walt, S.~J., {et~al.} 2020, Nature, 585,
  357–362

\bibitem[{{Hartigan} {et~al.}(1989){Hartigan}, {Hartmann}, {Kenyon}, {Hewett},
  \& {Stauffer}}]{hartigan89}
{Hartigan}, P., {Hartmann}, L., {Kenyon}, S., {Hewett}, R., \& {Stauffer}, J.
  1989, \apjs, 70, 899

\bibitem[{{Hartmann} {et~al.}(1986){Hartmann}, {Hewett}, {Stahler}, \&
  {Mathieu}}]{hartmann86}
{Hartmann}, L., {Hewett}, R., {Stahler}, S., \& {Mathieu}, R.~D. 1986, \apj,
  309, 275

\bibitem[{{Hartmann} \& {Kenyon}(1990)}]{hartmann90}
{Hartmann}, L.~W., \& {Kenyon}, S.~J. 1990, \apj, 349, 190

\bibitem[{{Hayashi}(1961)}]{hayashi61}
{Hayashi}, C. 1961, \pasj, 13, 450

\bibitem[{{Herbig}(1962)}]{herbig62}
{Herbig}, G.~H. 1962, Advances in Astronomy and Astrophysics, 1, 47

\bibitem[{Herbig \& Bell(1988)}]{herbig88}
Herbig, G.~H., \& Bell, K.~R. 1988, in Third Catalog of Emission-Line Stars of
  the Orion Population : 3 : 1988

\bibitem[{{Herbst} {et~al.}(2007){Herbst}, {Eisl{\"o}ffel}, {Mundt}, \&
  {Scholz}}]{herbst07}
{Herbst}, W., {Eisl{\"o}ffel}, J., {Mundt}, R., \& {Scholz}, A. 2007, in
  Protostars and Planets V, ed. B.~{Reipurth}, D.~{Jewitt}, \& K.~{Keil}, 297

\bibitem[{{Herczeg} \& {Hillenbrand}(2014)}]{herczeg14}
{Herczeg}, G.~J., \& {Hillenbrand}, L.~A. 2014, The Astrophysical Journal, 786,
  97

\bibitem[{{Herczeg} \& {Hillenbrand}(2015)}]{herczeg15}
---. 2015, \apj, 808, 23

\bibitem[{{Hern{\'a}ndez} {et~al.}(2007){Hern{\'a}ndez}, {Calvet},
  {Brice{\~n}o}, {Hartmann}, {Vivas}, {Muzerolle}, {Downes}, {Allen}, \&
  {Gutermuth}}]{hernandez07}
{Hern{\'a}ndez}, J., {Calvet}, N., {Brice{\~n}o}, C., {et~al.} 2007, \apj, 671,
  1784

\bibitem[{{Huang} {et~al.}(2018){Huang}, {Andrews}, {Cleeves}, {{\"O}berg},
  {Wilner}, {Bai}, {Birnstiel}, {Carpenter}, {Hughes}, {Isella}, {P{\'e}rez},
  {Ricci}, \& {Zhu}}]{huang18}
{Huang}, J., {Andrews}, S.~M., {Cleeves}, L.~I., {et~al.} 2018, \apj, 852, 122

\bibitem[{Hunter(2007)}]{matplotlib}
Hunter, J.~D. 2007, Computing in Science Engineering, 9, 90

\bibitem[{{Hussaini} {et~al.}(2020){Hussaini}, {Mace}, {L{\'o}pez-Valdivia},
  {Honaker}, \& {Han}}]{hussaini20}
{Hussaini}, M., {Mace}, G.~N., {L{\'o}pez-Valdivia}, R., {Honaker}, E.~J., \&
  {Han}, E. 2020, Research Notes of the American Astronomical Society, 4, 241

\bibitem[{{Ireland} \& {Kraus}(2008)}]{ireland08}
{Ireland}, M.~J., \& {Kraus}, A.~L. 2008, \apjl, 678, L59

\bibitem[{Jaffe {et~al.}(1998)Jaffe, Keller, \& Ershov}]{jaffe98}
Jaffe, D.~T., Keller, L.~D., \& Ershov, O.~A. 1998, in Infrared Astronomical
  Instrumentation, ed. A.~M. Fowler, Vol. 3354, International Society for
  Optics and Photonics (SPIE), 201 -- 212.
\newblock \url{https://doi.org/10.1117/12.317263}

\bibitem[{{Johns-Krull}(2007)}]{johns-krull07}
{Johns-Krull}, C.~M. 2007, \apj, 664, 975

\bibitem[{{Johns-Krull} \& {Valenti}(2001)}]{johns-krull01}
{Johns-Krull}, C.~M., \& {Valenti}, J.~A. 2001, \apj, 561, 1060

\bibitem[{{Johns-Krull} {et~al.}(1999){Johns-Krull}, {Valenti}, \&
  {Koresko}}]{johns-krull99}
{Johns-Krull}, C.~M., {Valenti}, J.~A., \& {Koresko}, C. 1999, \apj, 516, 900

\bibitem[{{Joy}(1949)}]{joy49}
{Joy}, A.~H. 1949, \apj, 110, 424

\bibitem[{{Kastner} {et~al.}(1997){Kastner}, {Zuckerman}, {Weintraub}, \&
  {Forveille}}]{katsner97}
{Kastner}, J.~H., {Zuckerman}, B., {Weintraub}, D.~A., \& {Forveille}, T. 1997,
  Science, 277, 67

\bibitem[{{Kenyon} {et~al.}(1994){Kenyon}, {Gomez}, {Marzke}, \&
  {Hartmann}}]{kenyon94}
{Kenyon}, S.~J., {Gomez}, M., {Marzke}, R.~O., \& {Hartmann}, L. 1994, \aj,
  108, 251

\bibitem[{{Kenyon} \& {Hartmann}(1987)}]{kenyon87}
{Kenyon}, S.~J., \& {Hartmann}, L. 1987, \apj, 323, 714

\bibitem[{{Kenyon} \& {Hartmann}(1995)}]{kenyon95}
---. 1995, \apjs, 101, 117

\bibitem[{{Koenigl}(1991)}]{koenigl91}
{Koenigl}, A. 1991, \apjl, 370, L39

\bibitem[{{Konopacky} {et~al.}(2007){Konopacky}, {Ghez}, {Rice}, \&
  {Duch{\^e}ne}}]{konopacky07}
{Konopacky}, Q.~M., {Ghez}, A.~M., {Rice}, E.~L., \& {Duch{\^e}ne}, G. 2007,
  \apj, 663, 394

\bibitem[{{Kraus} {et~al.}(2017){Kraus}, {Herczeg}, {Rizzuto}, {Mann},
  {Slesnick}, {Carpenter}, {Hillenbrand}, \& {Mamajek}}]{kraus17}
{Kraus}, A.~L., {Herczeg}, G.~J., {Rizzuto}, A.~C., {et~al.} 2017, \apj, 838,
  150

\bibitem[{{Kraus} \& {Hillenbrand}(2009)}]{kraus09}
{Kraus}, A.~L., \& {Hillenbrand}, L.~A. 2009, \apj, 704, 531

\bibitem[{{Kraus} {et~al.}(2011){Kraus}, {Ireland}, {Martinache}, \&
  {Hillenbrand}}]{kraus11}
{Kraus}, A.~L., {Ireland}, M.~J., {Martinache}, F., \& {Hillenbrand}, L.~A.
  2011, \apj, 731, 8

\bibitem[{{Kraus} {et~al.}(2006){Kraus}, {White}, \& {Hillenbrand}}]{kraus06}
{Kraus}, A.~L., {White}, R.~J., \& {Hillenbrand}, L.~A. 2006, \apj, 649, 306

\bibitem[{{Krolikowski} {et~al.}(2021){Krolikowski}, {Kraus}, \&
  {Rizzuto}}]{Krolikowski2021}
{Krolikowski}, D.~M., {Kraus}, A.~L., \& {Rizzuto}, A.~C. 2021, arXiv e-prints,
  arXiv:2105.13370

\bibitem[{{Lada} {et~al.}(2006){Lada}, {Muench}, {Luhman}, {Allen}, {Hartmann},
  {Megeath}, {Myers}, {Fazio}, {Wood}, {Muzerolle}, {Rieke}, {Siegler}, \&
  {Young}}]{lada06}
{Lada}, C.~J., {Muench}, A.~A., {Luhman}, K.~L., {et~al.} 2006, \aj, 131, 1574

\bibitem[{{Langer} \& {Kudritzki}(2014)}]{langer14}
{Langer}, N., \& {Kudritzki}, R.~P. 2014, \aap, 564, A52

\bibitem[{{Lavail} {et~al.}(2019){Lavail}, {Kochukhov}, \&
  {Hussain}}]{lavail19}
{Lavail}, A., {Kochukhov}, O., \& {Hussain}, G.~A.~J. 2019, \aap, 630, A99

\bibitem[{{Lavail} {et~al.}(2017){Lavail}, {Kochukhov}, {Hussain}, {Alecian},
  {Herczeg}, \& {Johns-Krull}}]{lavail17}
{Lavail}, A., {Kochukhov}, O., {Hussain}, G.~A.~J., {et~al.} 2017, \aap, 608,
  A77

\bibitem[{{Lee} {et~al.}(2017){Lee}, {Gullikson}, \& Kaplan}]{lee17}
{Lee}, J.-J., {Gullikson}, K., \& Kaplan, K. 2017, igrins/plp 2.2.0. Zenodo,
  doi:10.5281/zenodo.845059.
\newblock \url{https://doi.org/10.5281/zenodo.845059}

\bibitem[{{Leinert} {et~al.}(1991){Leinert}, {Haas}, {Mundt}, {Richichi}, \&
  {Zinnecker}}]{leinert91}
{Leinert}, C., {Haas}, M., {Mundt}, R., {Richichi}, A., \& {Zinnecker}, H.
  1991, \aap, 250, 407

\bibitem[{{Leinert} {et~al.}(1993){Leinert}, {Zinnecker}, {Weitzel},
  {Christou}, {Ridgway}, {Jameson}, {Haas}, \& {Lenzen}}]{leinert93}
{Leinert}, C., {Zinnecker}, H., {Weitzel}, N., {et~al.} 1993, \aap, 278, 129

\bibitem[{{Long} {et~al.}(2019){Long}, {Herczeg}, {Harsono}, {Pinilla},
  {Tazzari}, {Manara}, {Pascucci}, {Cabrit}, {Nisini}, {Johnstone}, {Edwards},
  {Salyk}, {Menard}, {Lodato}, {Boehler}, {Mace}, {Liu}, {Mulders}, {Hendler},
  {Ragusa}, {Fischer}, {Banzatti}, {Rigliaco}, {van de Plas}, {Dipierro},
  {Gully-Santiago}, \& {Lopez-Valdivia}}]{long19}
{Long}, F., {Herczeg}, G.~J., {Harsono}, D., {et~al.} 2019, \apj, 882, 49

\bibitem[{{L{\'o}pez-Santiago} {et~al.}(2016){L{\'o}pez-Santiago},
  {Crespo-Chac{\'o}n}, {Flaccomio}, {Sciortino}, {Micela}, \&
  {Reale}}]{lopez16}
{L{\'o}pez-Santiago}, J., {Crespo-Chac{\'o}n}, I., {Flaccomio}, E., {et~al.}
  2016, \aap, 590, A7

\bibitem[{L\'opez-Valdivia(2021{\natexlab{a}})}]{lopez-tau21}
L\'opez-Valdivia, R. 2021{\natexlab{a}}, {Reduced IGRINS Spectra:
  Taurus-Auriga}, vDRAFT VERSION,  Harvard Dataverse, doi:10.7910/DVN/F9N2E0.
\newblock \url{https://doi.org/10.7910/DVN/F9N2E0}

\bibitem[{L\'opez-Valdivia(2021{\natexlab{b}})}]{lopez-twa21}
---. 2021{\natexlab{b}}, {Reduced IGRINS spectra: TW Hydrae Association},
  vDRAFT VERSION,  Harvard Dataverse, doi:10.7910/DVN/BHSMT6.
\newblock \url{https://doi.org/10.7910/DVN/BHSMT6}

\bibitem[{L\'opez-Valdivia(2021{\natexlab{c}})}]{lopez-grid21}
---. 2021{\natexlab{c}}, {Codes and tools for IGRINS data: Synthetic grid},
  vDRAFT VERSION,  Harvard Dataverse, doi:10.7910/DVN/ALOPIS.
\newblock \url{https://doi.org/10.7910/DVN/ALOPIS}

\bibitem[{{L{\'o}pez-Valdivia} {et~al.}(2019){L{\'o}pez-Valdivia}, {Mace},
  {Sokal}, {Hussaini}, {Kidder}, {Mann}, {Gosnell}, {Oh}, {Kesseli},
  {Muirhead}, {Johns-Krull}, \& {Jaffe}}]{lopezval19}
{L{\'o}pez-Valdivia}, R., {Mace}, G.~N., {Sokal}, K.~R., {et~al.} 2019, \apj,
  879, 105

\bibitem[{{Luhman}(1999)}]{luhman99}
{Luhman}, K.~L. 1999, The Astrophysical Journal, 525, 466

\bibitem[{{Luhman}(2018)}]{luhman2018}
---. 2018, \aj, 156, 271

\bibitem[{{Luhman} {et~al.}(2010){Luhman}, {Allen}, {Espaillat}, {Hartmann}, \&
  {Calvet}}]{luhman2010}
{Luhman}, K.~L., {Allen}, P.~R., {Espaillat}, C., {Hartmann}, L., \& {Calvet},
  N. 2010, The Astrophysical Journal Supplement Series, 186, 111

\bibitem[{{Luhman} {et~al.}(2003{\natexlab{a}}){Luhman}, {Brice{\~n}o},
  {Stauffer}, {Hartmann}, {Barrado y Navascu{\'e}s}, \& {Caldwell}}]{luhman03}
{Luhman}, K.~L., {Brice{\~n}o}, C., {Stauffer}, J.~R., {et~al.}
  2003{\natexlab{a}}, The Astrophysical Journal, 590, 348

\bibitem[{{Luhman} {et~al.}(2017){Luhman}, {Mamajek}, {Shukla}, \&
  {Loutrel}}]{luhman17}
{Luhman}, K.~L., {Mamajek}, E.~E., {Shukla}, S.~J., \& {Loutrel}, N.~P. 2017,
  \aj, 153, 46

\bibitem[{{Luhman} {et~al.}(2003{\natexlab{b}}){Luhman}, {Stauffer}, {Muench},
  {Rieke}, {Lada}, {Bouvier}, \& {Lada}}]{luhman03b}
{Luhman}, K.~L., {Stauffer}, J.~R., {Muench}, A.~A., {et~al.}
  2003{\natexlab{b}}, The Astrophysical Journal, 593, 1093

\bibitem[{{Mace} {et~al.}(2016){Mace}, {Kim}, {Jaffe}, {Park}, {Lee}, {Kaplan},
  {Yu}, {Yuk}, {Chun}, {Pak}, {Kim}, {Lee}, {Sneden}, {Afsar}, {Pavel}, {Lee},
  {Oh}, {Jeong}, {Park}, {Kidder}, {Lee}, {Nguyen Le}, {McLane},
  {Gully-Santiago}, {Oh}, {Lee}, {Hwang}, \& {Park}}]{mace16}
{Mace}, G., {Kim}, H., {Jaffe}, D.~T., {et~al.} 2016, in \procspie, Vol. 9908,
  Society of Photo-Optical Instrumentation Engineers (SPIE) Conference Series,
  99080C

\bibitem[{{Mace} {et~al.}(2018){Mace}, {Sokal}, {Lee}, {Oh}, {Park}, {Lee},
  {Good}, {MacQueen}, {Oh}, {Kaplan}, {Kidder}, {Chun}, {Yuk}, {Jeong}, {Pak},
  {Kim}, {Nah}, {Lee}, {Yu}, {Hwang}, {Park}, {Kim}, {Chinn}, {Peck}, {Diaz},
  {Rutten}, {Prato}, {Jacoby}, {Cornelius}, {Hardesty}, {DeGroff}, {Dunham},
  {Levine}, {Nofi}, {Lopez-Valdivia}, {Weinberger}, \& {Jaffe}}]{mace18}
{Mace}, G., {Sokal}, K., {Lee}, J.-J., {et~al.} 2018, in Society of
  Photo-Optical Instrumentation Engineers (SPIE) Conference Series, Vol. 10702,
  \procspie, 107020Q

\bibitem[{{Mamajek}(2009)}]{mamajek09}
{Mamajek}, E.~E. 2009, in American Institute of Physics Conference Series, Vol.
  1158, Exoplanets and Disks: Their Formation and Diversity, ed. T.~{Usuda},
  M.~{Tamura}, \& M.~{Ishii}, 3--10

\bibitem[{{Mann} {et~al.}(2015){Mann}, {Feiden}, {Gaidos}, {Boyajian}, \& {von
  Braun}}]{mann15}
{Mann}, A.~W., {Feiden}, G.~A., {Gaidos}, E., {Boyajian}, T., \& {von Braun},
  K. 2015, \apj, 804, 64

\bibitem[{{Mann} {et~al.}(2013){Mann}, {Gaidos}, \& {Ansdell}}]{mann13}
{Mann}, A.~W., {Gaidos}, E., \& {Ansdell}, M. 2013, \apj, 779, 188

\bibitem[{{Mathieu} {et~al.}(1991){Mathieu}, {Adams}, \& {Latham}}]{mathieu91}
{Mathieu}, R.~D., {Adams}, F.~C., \& {Latham}, D.~W. 1991, \aj, 101, 2184

\bibitem[{{Mathieu} {et~al.}(1996){Mathieu}, {Martin}, \&
  {Magazzu}}]{mathieu96}
{Mathieu}, R.~D., {Martin}, E.~L., \& {Magazzu}, A. 1996, in American
  Astronomical Society Meeting Abstracts, Vol. 188, American Astronomical
  Society Meeting Abstracts \#188, 60.05

\bibitem[{{Muzerolle} {et~al.}(2003){Muzerolle}, {Calvet}, {Hartmann}, \&
  {D'Alessio}}]{muzerolle03}
{Muzerolle}, J., {Calvet}, N., {Hartmann}, L., \& {D'Alessio}, P. 2003, \apjl,
  597, L149

\bibitem[{{Natta} {et~al.}(2001){Natta}, {Prusti}, {Neri}, {Wooden}, {Grinin},
  \& {Mannings}}]{natta01}
{Natta}, A., {Prusti}, T., {Neri}, R., {et~al.} 2001, \aap, 371, 186

\bibitem[{{Newton} {et~al.}(2015){Newton}, {Charbonneau}, {Irwin}, \&
  {Mann}}]{newton15}
{Newton}, E.~R., {Charbonneau}, D., {Irwin}, J., \& {Mann}, A.~W. 2015, \apj,
  800, 85

\bibitem[{{Nguyen} {et~al.}(2012){Nguyen}, {Brandeker}, {van Kerkwijk}, \&
  {Jayawardhana}}]{nguyen12}
{Nguyen}, D.~C., {Brandeker}, A., {van Kerkwijk}, M.~H., \& {Jayawardhana}, R.
  2012, \apj, 745, 119

\bibitem[{{Nguyen} {et~al.}(2009){Nguyen}, {Jayawardhana}, {van Kerkwijk},
  {Brandeker}, {Scholz}, \& {Damjanov}}]{nguyen09}
{Nguyen}, D.~C., {Jayawardhana}, R., {van Kerkwijk}, M.~H., {et~al.} 2009,
  \apj, 695, 1648

\bibitem[{{Nofi} {et~al.}(2021){Nofi}, {Johns-Krull}, {L{\'o}pez-Valdivia},
  {Biddle}, {Carvalho}, {Huber}, {Jaffe}, {Llama}, {Mace}, {Prato}, {Skiff},
  {Sokal}, {Sullivan}, \& {Tayar}}]{nofi21}
{Nofi}, L.~A., {Johns-Krull}, C.~M., {L{\'o}pez-Valdivia}, R., {et~al.} 2021,
  arXiv e-prints, arXiv:2103.01246

\bibitem[{{Padgett}(1996)}]{padgett96}
{Padgett}, D.~L. 1996, \apj, 471, 847

\bibitem[{{Park} {et~al.}(2014){Park}, {Jaffe}, {Yuk}, {Chun}, {Pak}, {Kim},
  {Pavel}, {Lee}, {Oh}, {Jeong}, {Sim}, {Lee}, {Nguyen Le}, {Strubhar},
  {Gully-Santiago}, {Oh}, {Cha}, {Moon}, {Park}, {Brooks}, {Ko}, {Han}, {Nah},
  {Hill}, {Lee}, {Barnes}, {Yu}, {Kaplan}, {Mace}, {Kim}, {Lee}, {Hwang}, \&
  {Park}}]{park14}
{Park}, C., {Jaffe}, D.~T., {Yuk}, I.-S., {et~al.} 2014, in Society of
  Photo-Optical Instrumentation Engineers (SPIE) Conference Series, Vol. 9147,
  Society of Photo-Optical Instrumentation Engineers (SPIE) Conference Series,
  1

\bibitem[{{Pecaut} \& {Mamajek}(2013)}]{pecaut13}
{Pecaut}, M.~J., \& {Mamajek}, E.~E. 2013, \apjs, 208, 9

\bibitem[{{P{\'e}rez} {et~al.}(2020){P{\'e}rez}, {Casassus}, {Hales}, {Marino},
  {Cheetham}, {Zurlo}, {Cieza}, {Dong}, {Alarc{\'o}n}, {Ben{\'\i}tez-Llambay},
  {Fomalont}, \& {Avenhaus}}]{perez20}
{P{\'e}rez}, S., {Casassus}, S., {Hales}, A., {et~al.} 2020, \apjl, 889, L24

\bibitem[{{Rajpurohit} {et~al.}(2018){Rajpurohit}, {Allard}, {Rajpurohit},
  {Sharma}, {Teixeira}, {Mousis}, \& {Kamlesh}}]{rajpurohit18}
{Rajpurohit}, A.~S., {Allard}, F., {Rajpurohit}, S., {et~al.} 2018, \aap, 620,
  A180

\bibitem[{{Rajpurohit} {et~al.}(2013){Rajpurohit}, {Reyl{\'e}}, {Allard},
  {Homeier}, {Schultheis}, {Bessell}, \& {Robin}}]{rajpurohit13}
{Rajpurohit}, A.~S., {Reyl{\'e}}, C., {Allard}, F., {et~al.} 2013, \aap, 556,
  A15

\bibitem[{Rayner {et~al.}(2016)Rayner, Tokunaga, Jaffe, Bonnet, Ching,
  Connelley, Kokubun, Lockhart, \& Warmbier}]{rayner16}
Rayner, J., Tokunaga, A., Jaffe, D., {et~al.} 2016, in Ground-based and
  Airborne Instrumentation for Astronomy VI, ed. C.~J. Evans, L.~Simard, \&
  H.~Takami, Vol. 9908, International Society for Optics and Photonics (SPIE),
  2401 -- 2417.
\newblock \url{https://doi.org/10.1117/12.2232064}

\bibitem[{{Rebull} {et~al.}(2010){Rebull}, {Padgett}, {McCabe}, {Hillenbrand},
  {Stapelfeldt}, {Noriega-Crespo}, {Carey}, {Brooke}, {Huard}, {Terebey},
  {Audard}, {Monin}, {Fukagawa}, {G{\"u}del}, {Knapp}, {Menard}, {Allen},
  {Angione}, {Baldovin-Saavedra}, {Bouvier}, {Briggs}, {Dougados}, {Evans},
  {Flagey}, {Guieu}, {Grosso}, {Glauser}, {Harvey}, {Hines}, {Latter},
  {Skinner}, {Strom}, {Tromp}, \& {Wolf}}]{rebull2010}
{Rebull}, L.~M., {Padgett}, D.~L., {McCabe}, C.-E., {et~al.} 2010, \apjs, 186,
  259

\bibitem[{{Reid} \& {Hawley}(2005)}]{reid05}
{Reid}, I.~N., \& {Hawley}, S.~L. 2005, {New light on dark stars : red dwarfs,
  low-mass stars, brown dwarfs} (Springer, Berlin, Heidelberg),
  doi:10.1007/3-540-27610-6

\bibitem[{{Ribas} {et~al.}(2015){Ribas}, {Bouy}, \& {Mer{\'\i}n}}]{ribas15}
{Ribas}, {\'A}., {Bouy}, H., \& {Mer{\'\i}n}, B. 2015, \aap, 576, A52

\bibitem[{{Ribas} {et~al.}(2014){Ribas}, {Mer{\'\i}n}, {Bouy}, \&
  {Maud}}]{ribas14}
{Ribas}, {\'A}., {Mer{\'\i}n}, B., {Bouy}, H., \& {Maud}, L.~T. 2014, \aap,
  561, A54

\bibitem[{{Richichi} {et~al.}(1999){Richichi}, {Ragland}, {Calamai}, {Baffa},
  {Stecklum}, \& {Richter}}]{richichi99}
{Richichi}, A., {Ragland}, S., {Calamai}, G., {et~al.} 1999, \aap, 350, 491

\bibitem[{{Riedel} {et~al.}(2014){Riedel}, {Finch}, {Henry}, {Subasavage},
  {Jao}, {Malo}, {Rodriguez}, {White}, {Gies}, {Dieterich}, {Winters},
  {Davison}, {Nelan}, {Blunt}, {Cruz}, {Rice}, \& {Ianna}}]{riedel14}
{Riedel}, A.~R., {Finch}, C.~T., {Henry}, T.~J., {et~al.} 2014, \aj, 147, 85

\bibitem[{{Ru{\'\i}z-Rodr{\'\i}guez} {et~al.}(2016){Ru{\'\i}z-Rodr{\'\i}guez},
  {Ireland}, {Cieza}, \& {Kraus}}]{ruiz16}
{Ru{\'\i}z-Rodr{\'\i}guez}, D., {Ireland}, M., {Cieza}, L., \& {Kraus}, A.
  2016, \mnras, 463, 3829

\bibitem[{{Ryabchikova} {et~al.}(2015){Ryabchikova}, {Piskunov}, {Kurucz},
  {Stempels}, {Heiter}, {Pakhomov}, \& {Barklem}}]{vald3}
{Ryabchikova}, T., {Piskunov}, N., {Kurucz}, R.~L., {et~al.} 2015, \physscr,
  90, 054005

\bibitem[{{Santos} {et~al.}(2008){Santos}, {Melo}, {James}, {Gameiro},
  {Bouvier}, \& {Gomes}}]{santos08}
{Santos}, N.~C., {Melo}, C., {James}, D.~J., {et~al.} 2008, \aap, 480, 889

\bibitem[{{Schaefer} {et~al.}(2014){Schaefer}, {Prato}, {Simon}, \&
  {Patience}}]{schaefer14}
{Schaefer}, G.~H., {Prato}, L., {Simon}, M., \& {Patience}, J. 2014, \aj, 147,
  157

\bibitem[{{Schmidt-Kaler}(1982)}]{schmidt82}
{Schmidt-Kaler}, T. 1982, in Landolt-B{\"o}rnstein - Group VI Astronomy and
  Astrophysics, ed. K.~Schaifers \& H.~H. Voigt, Vol.~2B (Springer-Verlag
  Berlin Heidelberg).
\newblock
  \url{https://materials.springer.com/lb/docs/sm_lbs_978-3-540-31378-6_1}

\bibitem[{{Simon} {et~al.}(1992){Simon}, {Chen}, {Howell}, {Benson}, \&
  {Slowik}}]{simon92}
{Simon}, M., {Chen}, W.~P., {Howell}, R.~R., {Benson}, J.~A., \& {Slowik}, D.
  1992, \apj, 384, 212

\bibitem[{{Simon} {et~al.}(2019){Simon}, {Guilloteau}, {Beck}, {Chapillon}, {Di
  Folco}, {Dutrey}, {Feiden}, {Grosso}, {Pi{\'e}tu}, {Prato}, \&
  {Schaefer}}]{simon19}
{Simon}, M., {Guilloteau}, S., {Beck}, T.~L., {et~al.} 2019, \apj, 884, 42

\bibitem[{{Sneden}(1973)}]{moog}
{Sneden}, C.~A. 1973, PhD thesis, THE UNIVERSITY OF TEXAS AT AUSTIN.

\bibitem[{{Soderblom} {et~al.}(1989){Soderblom}, {Pendleton}, \&
  {Pallavicini}}]{soderblom89}
{Soderblom}, D.~R., {Pendleton}, J., \& {Pallavicini}, R. 1989, \aj, 97, 539

\bibitem[{{Sokal} {et~al.}(2018){Sokal}, {Deen}, {Mace}, {Lee}, {Oh}, {Kim},
  {Kidder}, \& {Jaffe}}]{sokal18}
{Sokal}, K.~R., {Deen}, C.~P., {Mace}, G.~N., {et~al.} 2018, \apj, 853, 120

\bibitem[{{Sokal} {et~al.}(2020){Sokal}, {Johns-Krull}, {Mace}, {Nofi},
  {Prato}, {Lee}, \& {Jaffe}}]{sokal20}
{Sokal}, K.~R., {Johns-Krull}, C.~M., {Mace}, G.~N., {et~al.} 2020, \apj, 888,
  116

\bibitem[{{Strai{\v{z}}ys}(1992)}]{straivzys92}
{Strai{\v{z}}ys}, V. 1992, {Multicolor stellar photometry} (Pachart Publishing
  House)

\bibitem[{{Valenti} {et~al.}(1993){Valenti}, {Basri}, \& {Johns}}]{valenti93}
{Valenti}, J.~A., {Basri}, G., \& {Johns}, C.~M. 1993, \aj, 106, 2024

\bibitem[{{Varga} {et~al.}(2017){Varga}, {Gab{\'a}nyi}, {{\'A}brah{\'a}m},
  {Chen}, {K{\'o}sp{\'a}l}, {Menu}, {Ratzka}, {van Boekel}, {Dullemond},
  {Henning}, {Jaffe}, {Juh{\'a}sz}, {Mo{\'o}r}, {Mosoni}, \& {Sipos}}]{varga17}
{Varga}, J., {Gab{\'a}nyi}, K.~{\'E}., {{\'A}brah{\'a}m}, P., {et~al.} 2017,
  \aap, 604, A84

\bibitem[{Virtanen {et~al.}(2020)Virtanen, Gommers, Oliphant, Haberland, Reddy,
  Cournapeau, Burovski, Peterson, Weckesser, Bright, {van der Walt}, Brett,
  Wilson, Millman, Mayorov, Nelson, Jones, Kern, Larson, Carey, Polat, Feng,
  Moore, {VanderPlas}, Laxalde, Perktold, Cimrman, Henriksen, Quintero, Harris,
  Archibald, Ribeiro, Pedregosa, {van Mulbregt}, \& {SciPy 1.0
  Contributors}}]{scipy}
Virtanen, P., Gommers, R., Oliphant, T.~E., {et~al.} 2020, Nature Methods, 17,
  261

\bibitem[{{Walter} {et~al.}(1988){Walter}, {Brown}, {Mathieu}, {Myers}, \&
  {Vrba}}]{walter88}
{Walter}, F.~M., {Brown}, A., {Mathieu}, R.~D., {Myers}, P.~C., \& {Vrba},
  F.~J. 1988, \aj, 96, 297

\bibitem[{Wang {et~al.}(2010)Wang, Gully-Santiago, Deen, Mar, \&
  Jaffe}]{wang10}
Wang, W., Gully-Santiago, M., Deen, C., Mar, D.~J., \& Jaffe, D.~T. 2010, in
  Modern Technologies in Space- and Ground-based Telescopes and
  Instrumentation, ed. E.~Atad-Ettedgui \& D.~Lemke, Vol. 7739, International
  Society for Optics and Photonics (SPIE), 1575 -- 1583.
\newblock \url{https://doi.org/10.1117/12.857164}

\bibitem[{{Webb} {et~al.}(1999){Webb}, {Zuckerman}, {Platais}, {Patience},
  {White}, {Schwartz}, \& {McCarthy}}]{webb99}
{Webb}, R.~A., {Zuckerman}, B., {Platais}, I., {et~al.} 1999, \apjl, 512, L63

\bibitem[{{Weintraub}(1989)}]{weintraub89}
{Weintraub}, D.~A. 1989, PhD thesis, California Univ., Los Angeles.

\bibitem[{{W}es {M}c{K}inney(2010)}]{pandas}
{W}es {M}c{K}inney. 2010, in {P}roceedings of the 9th {P}ython in {S}cience
  {C}onference, ed. {S}t\'efan van~der {W}alt \& {J}arrod {M}illman, 56 -- 61

\bibitem[{{Wright} \& {Eastman}(2014)}]{wright2014}
{Wright}, J.~T., \& {Eastman}, J.~D. 2014, \pasp, 126, 838

\bibitem[{{Yang} \& {Johns-Krull}(2011)}]{yang11}
{Yang}, H., \& {Johns-Krull}, C.~M. 2011, \apj, 729, 83

\bibitem[{{Yang} {et~al.}(2005){Yang}, {Johns-Krull}, \& {Valenti}}]{yang05}
{Yang}, H., {Johns-Krull}, C.~M., \& {Valenti}, J.~A. 2005, \apj, 635, 466

\bibitem[{{Yao} {et~al.}(2018){Yao}, {Meyer}, {Covey}, {Tan}, \& {Da
  Rio}}]{yao18}
{Yao}, Y., {Meyer}, M.~R., {Covey}, K.~R., {Tan}, J.~C., \& {Da Rio}, N. 2018,
  \apj, 869, 72

\bibitem[{{Yuk} {et~al.}(2010){Yuk}, {Jaffe}, {Barnes}, {Chun}, {Park}, {Lee},
  {Lee}, {Wang}, {Park}, {Pak}, {Strubhar}, {Deen}, {Oh}, {Seo}, {Pyo}, {Park},
  {Lacy}, {Goertz}, {Rand}, \& {Gully-Santiago}}]{yuk10}
{Yuk}, I.-S., {Jaffe}, D.~T., {Barnes}, S., {et~al.} 2010, in \procspie, Vol.
  7735, Ground-based and Airborne Instrumentation for Astronomy III, 77351M

\bibitem[{{Zuckerman} \& {Song}(2004)}]{zuckerman04}
{Zuckerman}, B., \& {Song}, I. 2004, \araa, 42, 685

\end{thebibliography}

\end{document}